\documentclass[a4paper,11pt]{article}
\pdfoutput=1
\usepackage{tikz}
\usepackage[margin=2.0cm]{geometry}
\usepackage{amsmath,amssymb,bm}
\usepackage{jheparxiv}
\usepackage{physics}
\usepackage[compat=1.1.0]{tikz-feynhand}
\usetikzlibrary{patterns,decorations.pathreplacing}
\usepackage[T1]{fontenc}
\usepackage{here}
\usepackage{tcolorbox}
\usepackage{comment}
\usepackage{mathrsfs}
\usepackage{dsfont}
\usepackage{braket}
\usepackage{booktabs}
\tcbuselibrary{skins}
\newcommand{\nn}{\nonumber \\}

\newcommand{\W}{\mathcal W}

\newcommand{\Amp}{ \mathcal{A} }

\newcommand{\re}{{\rm Re}\,}
\newcommand{\ima}{{\rm Im}\,}
\title{
\shortstack[l]{
An Effective $S$-Matrix Approach to  \\[6pt]
Low-Frequency Waveforms from Black Hole Mergers}
}
\author[a]{\fontsize{10pt}{12pt}\selectfont
Katsuki Aoki,}
\affiliation[a]{Graduate School of Science and Engineering, Saitama University, 255 Shimo-Okubo, Sakura-ku, Saitama 338-8570, Japan}
\emailAdd{katsukiaoki@mail.saitama-u.ac.jp}
\author[b,c]{Feng-Yin Cheng,}
\affiliation[b]{Department of Physics, Columbia University, New York, NY 10027, USA}
\affiliation[c]{Department of Physics and Center for Theoretical Physics, National Taiwan University, Taipei
10617, Taiwan}
\emailAdd{fc2871@columbia.edu}
\author[d]{Andrea~Cristofoli,}
\affiliation[d]{Center for Gravitational Physics and Quantum Information, Yukawa Institute for Theoretical Physics, Kyoto University, 606-8502, Kyoto, Japan}
\emailAdd{cristofoli@yukawa.kyoto-u.ac.jp}
\author[c,e]{Yu-tin~Huang,}
\affiliation[e]{Physics Division, National Center for Theoretical Sciences, Taipei 10617, Taiwan}
\emailAdd{yutin@phys.ntu.edu.tw}
\author[f,g,h]{and~Hyun~Jeong}
\emailAdd{jeong\_hyun@resceu.s.u-tokyo.ac.jp}
\affiliation[f]{Kavli IPMU (WPI), UTIAS, The University of Tokyo, Kashiwa, Chiba 277-8583, Japan}
\affiliation[g]{Research Center for the Early Universe (RESCEU), Graduate School of Science, The University
of Tokyo, Tokyo 113-0033, Japan}
\affiliation[h]{Department of Physics, Graduate School of Science, The University of Tokyo, Tokyo 113-0033, Japan}

\abstract{We develop an on-shell description of low-frequency gravitational waveforms from black-hole mergers beyond leading order. Treating the strongly coupled merger as effective hard $S$-matrix data, we organize its long-wavelength response using soft theorems and the KMOC formalism. At next-to-leading order, the quantum soft theorem contains logarithmic terms absent from the classical soft theorem. We show that these extra terms cancel in the full KMOC in-in observable between the one-loop radiative amplitude and the corresponding graviton cut, leaving precisely the classical logarithmic contributions associated with gravitational drag and early-time acceleration. We also identify the $1/\omega$ corrections from remnant recoil and Christodoulou non-linear memory. These results reveal a hierarchy of merger information accessible at low frequency: logarithmic tails depend only on asymptotic hard data, recoil probes total radiated momentum, while non-linear memory probes the angular distribution of the emitted radiation.}

\tikzset{
  cutdiag/.style={
    baseline={([yshift=-0.6ex]current bounding box.center)},
    x=8mm, y=8mm, line join=round, line cap=round},
  sc/.style={line width=0.9pt},                                  
  ms/.style={line width=0.55pt, double, double distance=1.3pt},  
  gr/.style={line width=0.55pt, decorate,
             decoration={snake, amplitude=0.45mm, segment length=1.5mm}},
  blob/.style={circle, fill=gray!55, draw, line width=0.6pt,
               inner sep=0pt, minimum size=3.4mm},
  lab/.style={font=\scriptsize, inner sep=0pt},
}

\newcommand{\flavms}{ms}
\newcommand{\flavsc}{sc}
\newcommand{\flav}[1]{\if#1X\flavms\else\flavsc\fi}

\newcommand{\extleg}[4][sc]{%
  \draw[#1] (#2) -- ++(#3) coordinate (tmpE);
  \node[lab] at ($(tmpE)!-7pt!(#2)$) {#4};}

\NewDocumentCommand{\FiveTree}{O{} O{}}{%
\begin{tikzpicture}[cutdiag]
\coordinate (V) at (0,0);
\extleg[gr]{V}{-0.55,0.45}{$k^{#1}$}
\extleg[gr]{V}{-0.7,-0.05}{$\ell^{#2}$}
\extleg[sc]{V}{0.6,0.35}{$1$}
\extleg[sc]{V}{0.6,-0.35}{$2$}
\extleg[ms]{V}{-0.15,-0.65}{$X$}
\node[blob, minimum size=4.6mm] at (V) {};
\end{tikzpicture}}

\NewDocumentCommand{\ChA}{O{} m m m O{} O{}}{%
\begin{tikzpicture}[cutdiag,#1]
\coordinate (L) at (0,0.5); \coordinate (R) at (1.6,0.5);
\draw[\flav{#2}] (L) -- (R);
\node[lab] at ($($(L)!0.5!(R)$)+(0,0.3)$) {$\hat p_{#2k}$};
\extleg[gr]{R}{0.45,0.5}{$\hat k^{#5}$}
\extleg[\flav{#2}]{R}{0.45,-0.5}{$#2$}
\extleg[gr]{L}{-0.45,0.5}{$\hat\ell^{#6}$}
\extleg[\flav{#3}]{L}{-0.62,0}{$#3$}
\extleg[\flav{#4}]{L}{-0.45,-0.5}{$#4$}
\node[blob] at (L) {};
\end{tikzpicture}}

\NewDocumentCommand{\ChB}{O{} m m m O{} O{}}{%
\begin{tikzpicture}[cutdiag,#1]
\coordinate (L) at (0,0.5); \coordinate (R) at (1.6,0.5);
\draw[\flav{#2}] (L) -- (R);
\node[lab] at ($($(L)!0.5!(R)$)+(0,0.3)$) {$\hat p_{#2\ell}$};
\extleg[gr]{L}{-0.45,0.5}{$\hat\ell^{#6}$}
\extleg[\flav{#2}]{L}{-0.45,-0.5}{$#2$}
\extleg[gr]{R}{0.45,0.5}{$\hat k^{#5}$}
\extleg[\flav{#3}]{R}{0.62,0}{$#3$}
\extleg[\flav{#4}]{R}{0.45,-0.5}{$#4$}
\node[blob] at (R) {};
\end{tikzpicture}}

\newcommand{\hdelta}{\hat{\delta}}
\newcommand{\hd}{\hat{\dd}}
\newcommand{\intIR}{\int_{\text{IR-log}} \hspace{-15pt} \hd^4 \ell \, }

\newcommand{\Stree}[2]{S^{(#1)}_{#2,\text{tree}}}
\newcommand{\Sloop}[2]{S^{(#1)}_{#2,\text{1-loop}}}
\newcommand{\hStree}[2]{\hat{S}^{(#1)}_{#2,\text{tree}}}
\newcommand{\hSloop}[2]{\hat{S}^{(#1)}_{#2,\text{1-loop}}}

\begin{document}

\begin{flushright}
STUPP-26-303,
RESCEU-24/26,
IPMU26-0033
\end{flushright}

\maketitle

\section{Introduction and summary}
\label{sec:intro}

\begin{figure}[h!]
\begin{center}
    \includegraphics[width=0.55\textwidth]{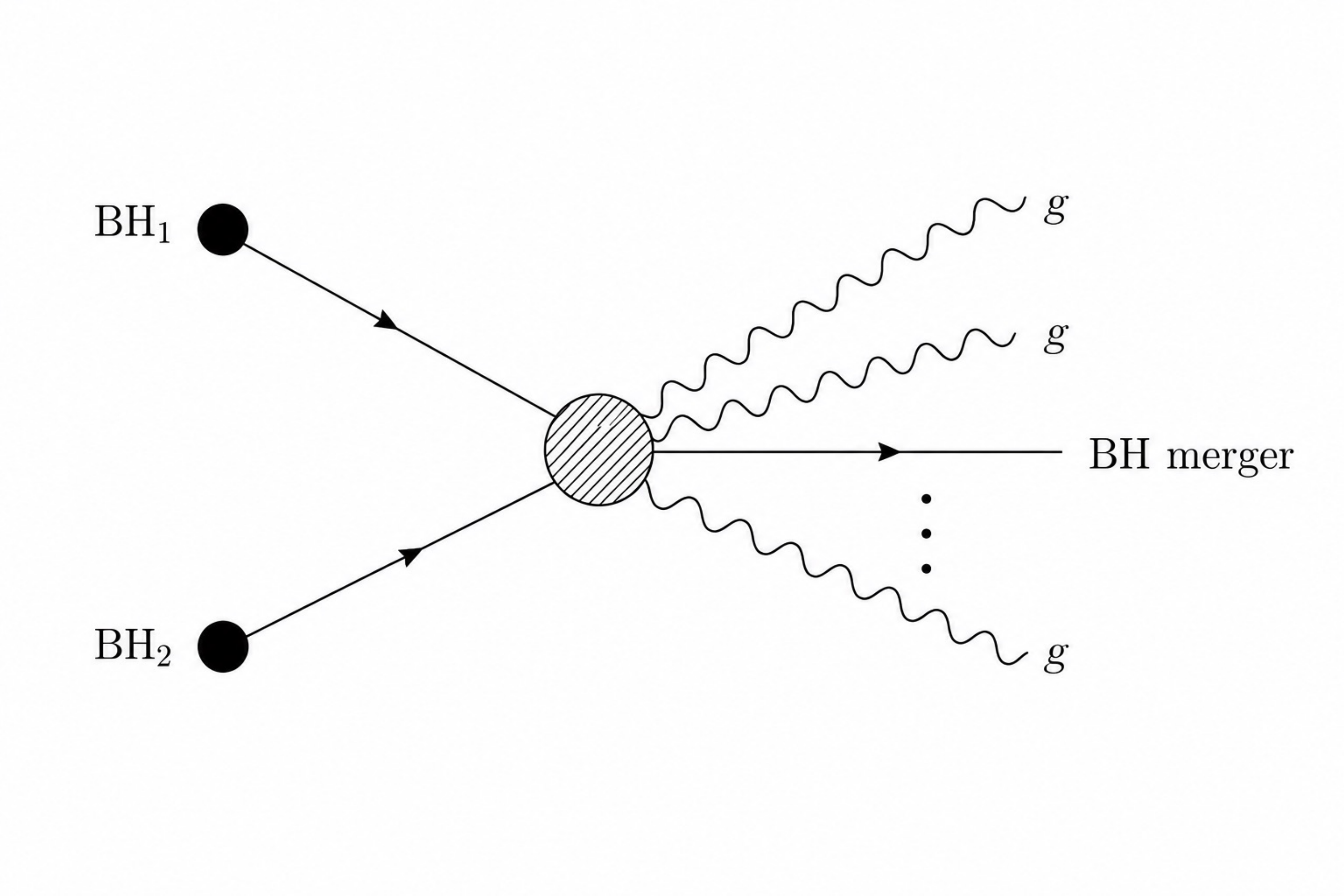}
\end{center}
\caption{We view a black hole merger as a fusion process of two massive particles into one accompanied by graviton emissions.}
\label{fig:BHmerger}
\end{figure}

Do black-hole mergers mark the ultimate limit of analytic predictivity?
In Ref.~\cite{Aoki:2024boe}, an amplitude-based framework was developed to address this question. The central idea in Ref.~\cite{Aoki:2024boe}
was to avoid resolving the strongly coupled coalescence and
instead describe the merger as an effective transition between asymptotic
black-hole states, as is common in particle physics; see Figure~\ref{fig:BHmerger} for illustration. This offers an effective description of black hole mergers and computes low-frequency parts of gravitational waves in an analytically controlled way.

In this formulation, utilising modern scattering-amplitude techniques, the merger dynamics is encoded in effective on-shell
data, while universal properties of the emitted radiation are extracted
without reconstructing the full strong-field evolution.\footnote{For earlier approaches valid also for the merger case but based
on Feynman diagrams or classical equations of motion, see
\cite{Sahoo:2018lxl,Saha:2019tub,Sahoo:2021ctw}.} The hard
\(\text{BH}_1+\text{BH}_2\to \text{BH}_X \) merger  transition is treated as genuinely non-perturbative input:
massive little-group covariance constrains its structure, while the
remaining spectral information encodes the unresolved strong-field
dynamics and is related to suitable inclusive black-hole observables. Embedding this construction into the KMOC formalism~\cite{Kosower:2018adc,Cristofoli:2021vyo}, Ref.~\cite{Aoki:2024boe} derived the leading low-frequency waveform and recovered the displacement memory to all orders in the classical angular momentum of the black-hole merger remnant.

A complete description of the low-frequency merger waveform, however, requires accounting for several effects beyond linear memory.\footnote{Soft constraints on classical observables have previously been explored within the KMOC formalism for inspiralling systems~\cite{Bautista:2021llr}. Our approach extends their applicability to merger events.} Once these effects are included, the waveform is no longer determined solely by the asymptotic massive states: one must also account for the dynamics of the radiation and its interplay with long-range gravitational interactions. These interactions generate logarithmic tails, while the energy and momentum carried away by the radiation induce the radiation reaction and set the recoil of the final black hole. Furthermore, the radiation itself acts as a gravitational source, giving rise to the non-linear Christodoulou memory.

Addressing these intertwined effects calls for a framework that can isolate their distinct contributions directly at the level of physical observables. Our approach meets this need by combining modern on-shell techniques, which avoid ambiguities associated with off-shell descriptions, with a systematic prescription for extracting the waveform produced by a merger process. In this way, it complements previous approaches and makes transparent how the various physical mechanisms contribute to the observable waveform.

Within this framework, we disentangle the different physical effects contributing to the waveform. In particular, we organize their contributions into two qualitatively distinct classes: universal contributions determined entirely by asymptotic data, and process-dependent contributions that probe the bulk dynamics of the merger. We elaborate on this distinction below.

\vspace{2mm}

\begin{figure}[h!]
\centering
\includegraphics[width=0.75\textwidth]{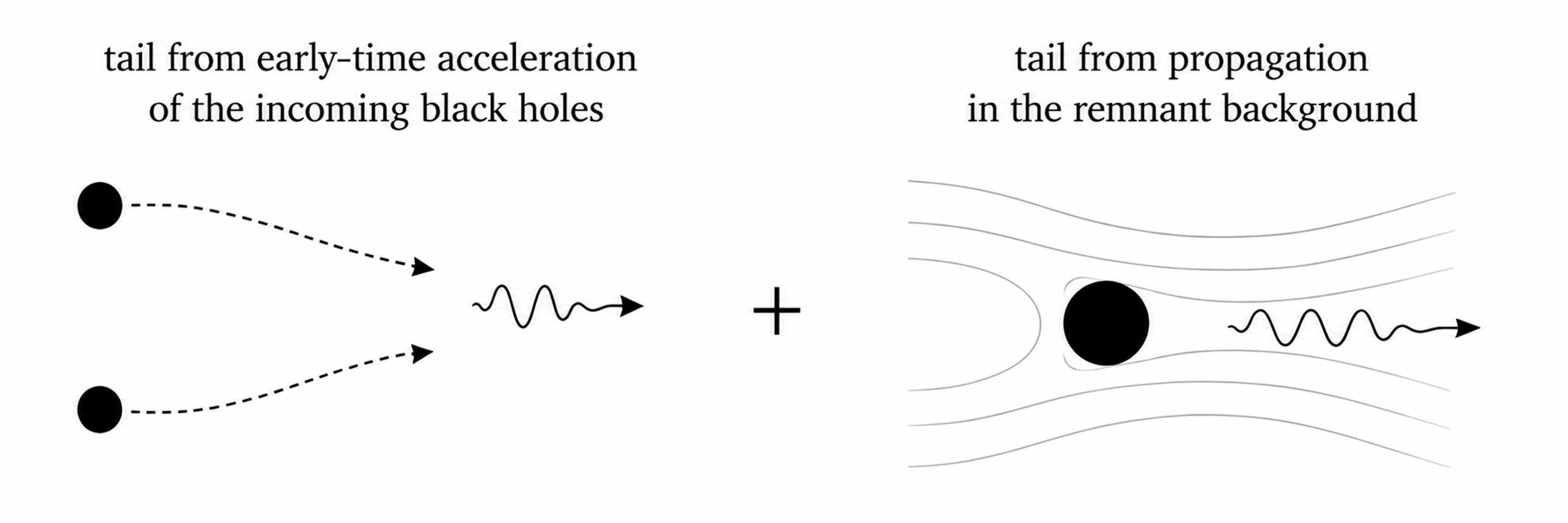}
\caption{Universal tails, represented by logarithms in the frequency space, originate from acceleration of the incoming black holes and propagation in the remnant gravitational field.}
    \label{fig:logarithmic_tails}
\end{figure}

\textbf{Universal waveform effects:} The first class arises from long-range gravitational interactions and manifests itself through logarithmic dependence on the observed frequency. In the time domain, these logarithms encode early-time or late-time power-law tails and are governed by the classical logarithmic soft theorem of Refs.~\cite{Sahoo:2018lxl,Saha:2019tub}. These tails have two distinct origins. One arises from radiation generated by the residual early-time acceleration of the incoming black holes, whereas the other results from the interaction of the emitted graviton with the long-range gravitational field of the final remnant. Both contributions are determined solely by the asymptotic hard-particle kinematics and are therefore insensitive to the details of the strong-field merger dynamics, as illustrated in Figure~\ref{fig:logarithmic_tails}.

\textbf{Non-universal waveform effects:} The second class comprises effects that are not determined by the asymptotic massive states alone, but instead depend on the radiation generated during the merger. The leading Weinberg pole, responsible for the linear memory effect, receives contributions from every finite-energy particle in the final state, including emitted gravitons. Consequently, the leading waveform is corrected both by the recoil of the final black hole and by the gravitational field sourced by the radiation itself, with the latter giving rise to the Christodoulou non-linear memory~\cite{Christodoulou:1991cr,Blanchet:1992br,Thorne:1992sdb}. Recent developments from the perspective of the on-shell program can be found in Refs.~\cite{Fernandes:2024xqr,Georgoudis:2025vkk,Kanno:2025how,Fernandes:2025double,Bini:2026vaq,Moynihan:2026xmv}. In contrast to the logarithmic tail, these effects therefore depend on inclusive information about the radiation produced during the merger and, through it, on the bulk dynamics of the merger process.

\textbf{Scope and strategy.}
Our aims are to systematically separate features of the low-frequency radiation which 
are fixed universally by soft physics and which require additional
information about the strong-field dynamics of the merger process. The latter will be packaged into effective amplitudes and can be determined by a finite number of strong-gravity inputs. We then formulate low-frequency merger waveforms including both universal and non-universal effects. 

The appropriate observable for making this separation is the KMOC
spectral waveform,
\begin{equation}
    i\mathcal W_\sigma(k)
    =
    \langle\Psi|
    S^\dagger \hat{a}_\sigma(k)S
    |\Psi\rangle .
    \label{eq:intro_KMOC}
\end{equation}
Crucially, $\mathcal W_\sigma$ is an in-in observable which can be expressed in terms of on-shell data after inserting a completeness relation of multi-particle states. Beyond linear memory, we find that the first non-trivial contributions are given by two terms in the kernel waveform which we can summarize as follows
\begin{equation}
i\W_\sigma=i\mathcal{W}^{(0)}_{\sigma} +i\mathcal{W}^{(1)}_{\sigma} + (n\geq 2)\,.
\end{equation}
The first two contributions are given by
\begin{equation}
\label{eq:first-terms}
i\W^{(0)}_\sigma(k)
= \int_X \widetilde\Gamma_3^*\,\widetilde\Gamma^\sigma_4(k)\,,
\qquad
i\W^{(1)}_\sigma(k)
= \sum_\eta \int_{X,\ell}
\bigl[\widetilde\Gamma^\eta_4(\ell)\bigr]^*
\widetilde\Gamma^{\sigma,\eta}_5(k,\ell)\,,
\end{equation}
and the on-shell transition amplitudes for the wavepackets are 
\begin{equation}
i\widetilde\Gamma^{\eta_1\cdots\eta_n}_{3+n}:=\bra{p_X,\alpha;\ell_1^{\eta_1};\cdots;\ell_n^{\eta_n}}S\ket{\Psi}\,.
\end{equation}
Here, $\ket{\Psi}$ is an initial wavepacket state describing the incoming black holes as in \cite{Kosower:2018adc}, $\bra{p_X,\alpha}$ denotes a formed (generically spinning) black hole, and $k$ and $\ell$ are observed and unobserved gravitons with helicities $\sigma$ and $\eta$. We also denote the frequency of $k$ by $\omega$. Let us explain the roles of the first term $\mathcal{W}^{(0)}$ and the second term with one graviton cut $\mathcal{W}^{(1)}$ in order.
\vspace{1mm}

\textbf{$\widetilde\Gamma_4$ at one-loop and the role of eikonal and massless poles:} The next-to-leading order (NLO) contribution to the waveform arising from the one-loop radiative amplitude is encoded in the first term of (\ref{eq:first-terms}) through $\widetilde\Gamma_4$. One-loop computations of gravitational waveforms have emerged as a fruitful arena for testing modern amplitude methods, particularly generalized-unitarity techniques, and for sharpening the connection between quantum-field-theoretic observables and classical gravitational radiation~\cite{Herderschee:2023subleading,Elkhidir:2023radiation,Georgoudis:2023inelastic,CaronHuot:2023asymptotic,Bini:2023comparison,Georgoudis:2023eke,Georgoudis:2024multipoles,Alessio:2024wmz,Bini:2024tale,Brunello:2025analytic,Bini:2026vaq,Brandhuber:2023oneLoop}. Our problem is, however, qualitatively different: rather than describing the radiation generated in an elastic scattering process, we must construct the one-loop radiative amplitude for an intrinsically inelastic transition in which two incoming black holes coalesce into a single remnant.
Restricting the explicit
calculation to the spin-independent sector, we construct the amplitude
using generalized unitarity and isolate the non-analytic terms that
control the logarithmic soft behaviour. \par

A useful feature of the
one-loop result is that its logarithms can be traced directly to the
poles of the reduced loop integrals\footnote{See \cite{Bern:2026oqp} for a recent similar application in the context of classical gravity from multi-loops.}. Two qualitatively different
classes appear. The eikonal poles reproduce precisely the logarithmic
structures predicted by the classical soft theorem: the configuration
involving the outgoing remnant and the observed graviton gives the gravitational drag, while the
incoming-pair configuration gives the tail generated by the long-range
acceleration of the incoming black holes. The massless graviton pole,
by contrast, produces additional logarithmic structures that are
present in the quantum soft theorem but have no counterpart in the
classical waveform. Crucially, these terms cannot be discarded on the basis of their $\hbar$ scaling alone, thereby demonstrating explicitly why the radiative amplitude alone is insufficient to recover the classical waveform.

\textbf{Cut graviton contributions and the role of double-soft and hierarchical regions:}
At the same perturbative order, the one-loop radiative contribution
must be supplemented, within the KMOC framework, by the one-graviton cut $\mathcal{W}^{(1)}_{\sigma}$, which involves an integration over the physical phase space of the unobserved graviton with momentum $\ell$. It is this cut integral---and not the one-loop
amplitude---that naturally separates into two parametrically distinct
regions,
\begin{equation}
    \ell\sim k\ll p_a,
    \qquad\qquad
    k\ll\ell \ll p_a,
    \label{eq:intro_regions}
\end{equation}
which we refer to as the double-soft and hierarchical regions,
respectively. In the double-soft region both the observed and cut
gravitons are soft, and the corresponding contribution provides the
inclusive completion of the logarithmic waveform. In particular, the
relevant Compton factorization channel cancels precisely the additional
logarithms generated by the massless pole of the one-loop amplitude,
while leaving the classical eikonal contributions unaffected. The
hierarchical region has a different physical role: here the observed
graviton is much softer than the radiation produced during the merger,
so that the latter becomes part of the hard final state. This region
does not participate in the cancellation of the quantum logarithms;
instead, it gives the recoil of the final black hole and the
Christodoulou non-linear memory, thereby introducing progressively more
inclusive information about the finite-frequency radiation generated
during the merger.

\textbf{Main results:} These analyses establish a prescription for describing waveforms of black-hole mergers at frequencies well below the characteristic or cutoff scale $\omega_*\sim 1/GM$. The separation of soft physics and strong-gravity physics is achieved by splitting unobserved cut-gravitons into soft $\ell \ll \omega_*$ and finite $\ell \sim \omega_*$, and a low-frequency waveform is obtained by combining the soft theorems and the KMOC formalism.

When both observed and unobserved radiative momenta are soft, this contribution is completely controlled by soft theorems and hence by the asymptotic hard data. This refines the relation between the classical low-frequency waveform and the quantum soft theorems. The classical limit of the quantum soft factor alone does not provide the classical waveform. The waveform requires to properly take multiple soft emissions into account, controlled by the double soft theorems~\cite{Cachazo:2015ksa,Saha:2016kjr,Saha:2017yqi,Chakrabarti:2017ltl}. Hence, the universal part of the low-frequency waveform can serve as a natural playground for not only making analytic predictions for black hole mergers but also developing deeper understanding of IR structure of gravity. In fact, our consistency check---the additional logarithmic terms arising from the one-loop quantum soft factor cancel against the corresponding double-soft cut---would be regarded as an example of the outcome arising from interplay between quantum soft theorems and classical physics. The cancellation shows that the real part of the one-loop soft factor, not the imaginary part as usually constrained by unitarity, also admits a certain factorisation structure into tree-level soft factors.

On the other hand, as for finite-frequency cut-gravitons, rather than attempting to resolve the strongly coupled merger dynamics itself, one replaces the hard process by an effective on-shell merger \(S\)-matrix for \(\text{BH}_1+\text{BH}_2\to \text{BH}_X+\{\ell_i\}\), while treating the observed graviton with frequency \(\omega\ll \omega_*\) through its universal soft coupling to the corresponding hard final state. The hard amplitudes contribute to low-frequency observables through particular integrations of the distribution of radiated states, requiring only a limited number of inclusive data supplied by the strong-gravity merger dynamics. At the order considered in this paper, only two data are required: the total radiated momentum determines the recoil of the remnant, while the angular energy distribution determines the Christodoulou memory. This leads to a systematic low-frequency description in which the unresolved strong-field region is encoded in a set of inclusive merger data, whereas their coupling to the measured long-wavelength radiation is fixed universally. The effective merger \(S\)-matrix therefore provides a factorized prescription: one first matches the hard merger onto the minimal set of inclusive observables resolved at a given order in the soft expansion, and then uses soft theorems, together with the KMOC formalism, to map these data into predictions for the low-frequency waveform.

\vspace{1mm}

The remainder of this paper is organized as follows. In Sec.~\ref{sec:Sensoft}, we review
the classical soft-graviton theorem and specialize its low-frequency
predictions to the merger case. In Sec.~\ref{sec:KMOC}, we formulate the
merger waveform as a KMOC in-in observable and derive its decomposition into contributions from the double-soft and hierarchical regions. We also clarify the relation between the quantum and classical soft theorems by adopting the wavepacket formulation. In Sec.~\ref{sec:oneloop}, we compute the logarithmic part of the one-loop
merger amplitude, separate its eikonal- and massless-pole
contributions, and evaluate the corresponding one-graviton cut. We
demonstrate the cancellation of spurious logarithms and recover
the universal classical logarithmic waveform. We then assemble the low-frequency merger waveform at the next-to-leading order, including the recoil and non-linear memory. We conclude in Sec.~\ref{sec:conclusion}.

\textbf{Conventions:} Unless stated otherwise, we work in natural units, $c=\hbar=1$, restoring $\hbar$ when discussing the classical limit, with the gravitational coupling $\kappa^2=32\pi G$. We adopt the mostly minus signature of the metric $\eta_{\mu\nu}={\rm diag}[+1,-1,-1,-1]$. The on-shell phase-space integral is denoted by
\begin{align}
    \int_p :=\int \dd \Phi(p)\,, \qquad \dd \Phi(p):=\hat{\dd}^4p \: \hat{\delta}^{(+)}(p^2-m^2)
\end{align}
with the $(2\pi)$ normalised measure and delta functions
\begin{align}
    \hat{\dd}^n p:= \frac{\dd^n p}{(2\pi)^n}\,, \qquad \hat{\delta}^{(n)}(p):=(2\pi)^n \delta^{(n)}(p)
    \,.
\end{align}
When the dimension of a delta function is clear from its argument, we suppress the superscript on $\hat{\delta}$.
Scattering amplitudes are collectively denoted by $\Amp$. Following Ref.~\cite{Aoki:2024boe}, we specify the scattering process through the arguments of $\Amp$, separating the out- and in-states by a vertical bar. In accordance with bra-ket notation, the labels to the left (right) of the vertical bar correspond to the out- (in-)states, respectively, while labels associated with different particles are separated by semicolons. When the arguments are suppressed, a subscript indicates the number of external legs.

We use $\sigma=\pm$ for the observed graviton helicity and $\eta=\pm$ for a cut-graviton helicity. Their polarization vectors are denoted by $\epsilon_k^\sigma$ and $\epsilon_\ell^\eta$, with polarization tensors $\varepsilon_{k,\mu\nu}^\sigma:=\epsilon_{k,\mu}^\sigma\epsilon_{k,\nu}^\sigma$ and $\varepsilon_{\ell,\mu\nu}^\eta:=\epsilon_{\ell,\mu}^\eta\epsilon_{\ell,\nu}^\eta$. Helicity labels are suppressed when no ambiguity arises. When presenting explicit expressions for scattering amplitudes, we employ the spinor-helicity conventions of Ref.~\cite{Arkani-Hamed:2017jhn}.

In this paper, we use several versions of soft theorems in different contexts. We summarize their notations as follows. First, the classical soft theorems are the statements of the low-frequency expansion of classical radiative gravitational field. Such soft waveforms are denoted as 
\begin{equation}
S_{\rm gr}(\varepsilon,k)\,.
\end{equation}
Second, the quantum soft theorems are relations between a transition amplitude with soft particles and a lower-point amplitude with soft particles removed. The soft factors are denoted as 
$\hat{S}_k$ and $\hat{S}_{k_1,k_2}$ representing single and double-soft factors, with the subscript denoting the soft momenta. These can be further expanded in $\kappa$ (loops) and $\omega$. Schematically, we have
\begin{align}
    \hat{S}_{k}&=\underbrace{\left[\hStree{-1}{k} + \hStree{0}{k}+\cdots\right]}_{\text{tree}} +\underbrace{\left[\hSloop{0}{k} +\cdots\right]}_{\text{1-loop}} + (\text{higher-loops})        
    \,, \nonumber\\
    \hat{S}_{k_1,k_2}&=\underbrace{\left[ \hStree{-2}{k_1,k_2} + \hStree{-1}{k_1,k_2}+\cdots \right]}_{\text{tree}} + (\text{loops})\,,
\end{align}
with the superscript denoting the power of $\omega$. Note that we have suppressed helicity labels. If necessary, we put the helicity label on the graviton momenta, e.g., $\hStree{-1}{k^{\sigma}}$. Finally, we will introduce a wavepacket version of the soft theorems, whose soft factors are denoted without hat, e.g., $\Stree{-1}{k}$. The relations among these soft theorems will be explained in Sec.~\ref{sec:KMOC}.

\section{Universal low-frequency waveforms}
\label{sec:Sensoft}
Before introducing an amplitude-based construction of the merger waveform, we first summarize the universal information in a merger event that follows from the classical soft theorems~\cite{Laddha:2018myi,Laddha:2018vbn,Sahoo:2018lxl,Saha:2019tub,Laddha:2019yaj,Sahoo:2021ctw,Sen:2024qzb,Krishna:2023fxg}.  This perspective is useful because the classical soft theorem is formulated directly in terms of the asymptotic scattering data and therefore makes no reference to intermediate dynamics. In particular, it gives a set of low-frequency targets that any microscopic description of the merger radiation must reproduce.

\subsection{Classical soft graviton theorem in four dimensions}
\label{sec:SenGeneral}

Following Refs.~\cite{Laddha:2018myi,Laddha:2018vbn,Sahoo:2018lxl,Saha:2019tub,Laddha:2019yaj,Sahoo:2020ryf,Sahoo:2021ctw,Sen:2024qzb,Krishna:2023fxg}, we define the radiative gravitational field by
\begin{equation}
h_{\mu\nu}:=\frac12\left(g_{\mu\nu}-\eta_{\mu\nu}\right),
 \qquad
 e_{\mu\nu}:=h_{\mu\nu}-\frac12\eta_{\mu\nu}h^\rho{}_{\rho}\,.
 \label{eq:SenTraceReverse}
\end{equation}
We write the position of the detector as $\bm{x} = R\,\hat{\bm{n}}$, where $\hat{\bm{n}}$ is a unit vector specifying the direction of observation. At future null infinity, $R \to \infty$ with the retarded time $u = t - R$ and $\hat{\bm{n}}$ fixed, the classical soft factor in 3+1 dimensions is related to the frequency-space radiative field by

\begin{equation}
 \int \dd u\,e^{i\omega u}\,
 \varepsilon^{\mu\nu}e_{\mu\nu}(u,R\hat{\bm n})
 =-\frac{i}{4\pi R}\frac{\kappa}{2}\,S_{\rm gr}(\varepsilon,k),
 \qquad
 k^\mu=-\omega(1,\hat{\bm n})\,.
 \label{eq:SenSoftDefinition}
\end{equation}
The minus sign in $k^\mu$ reflects the all-incoming momentum convention used for the emitted graviton.  Refs.~\cite{Laddha:2018myi,Laddha:2018vbn,Sahoo:2018lxl,Saha:2019tub,Laddha:2019yaj,Sahoo:2020ryf,Sahoo:2021ctw,Sen:2024qzb} use mostly-plus signature and units $8\pi G=1$.  In the equations below, we have translated their result to the mostly-minus signature used in this paper and restored $\kappa^2=32\pi G$. The momentum of a physically outgoing hard particle is written with an additional minus sign, and hence we introduce
\begin{equation}
 \varsigma_a=+1\quad\hbox{for an incoming hard leg},
 \qquad
 \varsigma_a=-1\quad\hbox{for an outgoing hard leg}.
 \label{eq:varsigmaDef}
\end{equation}
For a graviton of helicity $\sigma$ we write
\begin{equation}
 \varepsilon^{\sigma}_{k,\mu\nu}
 :=\epsilon^\sigma_{k,\mu}\epsilon^\sigma_{k,\nu},
 \qquad k\cdot\epsilon_k^\sigma=0,
 \label{eq:SenPolarization}
\end{equation}
and define the dimensionless null direction
\begin{equation}
 \bar k^\mu:=-\frac{k^\mu}{\omega}=(1,\hat{\bm n})
 \label{eq:nullDirection}
\end{equation}
together with
\begin{equation}
 D_{ab}:=\sqrt{(p_a\!\cdot p_b)^2-m_a^2m_b^2}\,.
 \label{eq:Dab}
\end{equation}

For a final state containing only massive hard particles, the universal classical terms of order $\omega^{-1}$ and $\ln\omega$ in Eq.~(2.6) of Ref.~\cite{Sahoo:2018lxl} take the form
\begin{align}
 S_{\rm gr}(\varepsilon,k)\Big|_{\omega^{-1},\ln\omega}
 ={}&
 - \frac{\kappa}{2}
 \sum_a\frac{\varepsilon_{\mu\nu}p_a^\mu p_a^\nu}{p_a\cdot k}
 \nonumber\\
 &-\frac{i\kappa^3}{32\pi}
 \bigl(\ln \omega ^{-1}+\ln R^{-1}\bigr)
 \left[\sum_{b:\,\varsigma_b=-1}k\cdot p_b\right]
 \left[\sum_a\frac{\varepsilon_{\mu\nu}p_a^\mu p_a^\nu}{p_a\cdot k}\right]
 \nonumber\\
 &+\frac{i\kappa^3}{64\pi}\ln\omega^{-1}
 \sum_a\frac{\varepsilon_{\mu\nu}p_a^\nu k_\rho}{p_a\cdot k}
 \sum_{\substack{b\neq a\\ \varsigma_a\varsigma_b=1}}
 \frac{p_a\cdot p_b}{D_{ab}^{3}}
 (p_b^\rho p_a^\mu-p_b^\mu p_a^\rho)
 \bigl[2(p_a\cdot p_b)^2-3m_a^2m_b^2\bigr] .
 \label{eq:SenClassical}
\end{align}
Only the non-analytic terms relevant below are displayed. The first line is the leading Weinberg term and gives the ordinary displacement memory~\cite{Weinberg:1965nx}.  The second line is the propagation, or gravitational-drag, contribution: the emitted graviton itself propagates through the long-range field sourced by the outgoing state. The third line is generated by the residual early-time ($\varsigma_a=+1$) or late-time ($\varsigma_a=-1$) acceleration of hard particles in their mutual long-range gravitational field. The logarithmic result was derived from the classical limit of the soft theorem in Ref.~\cite{Sahoo:2018lxl} and subsequently proved directly from the classical equations of motion in Ref.~\cite{Saha:2019tub}. Note that the argument of the logarithms must be understood as $(\omega+i\varepsilon)^{-1}$ for the gravitational drag and the late-time acceleration, while $(\omega-i \varepsilon)^{-1}$ for the early-time acceleration: in the time domain, they contribute to the late time $u \to \infty$ and early time $u\to -\infty$ waveforms, respectively. In the following, to reduce clutter, we suppress $i\varepsilon$ unless confusion arises whether $\ln \omega^{-1}$ contributes to early time or late time.

Two features of Eq.~\eqref{eq:SenClassical} will be important later. First, although the first line has been written above for massive hard particles, the Weinberg soft theorem applies to every finite-energy particle in the asymptotic state. Hence, at NLO, where backreactions from leading radiative effects should be taken into account, the physical final state is a sum of the recoiling remnant momentum $p_{X}$ and finite-energy emitted gravitons. Expanding around the leading remnant momentum produces the recoil contribution, while the Weinberg factors of the emitted gravitons give the non-linear, or null memory. Upon averaging over radiative final states, the latter is controlled by the graviton radiation spectrum. Second, the $\ln\omega$ terms displayed explicitly in the second and third lines of Eq.~\eqref{eq:SenClassical} are genuine logarithmic corrections, in contrast to the first line. Their classical coefficient can be rewritten without explicit dependence on the outgoing finite-energy massless particles, so that it is determined entirely by the incoming data and the outgoing massive states. Taken together, these observations show that at NLO, the $1/\omega$ coefficient requires information on the radiation spectrum, whereas the $\ln\omega$ coefficients are predicted from the hard asymptotic data alone.

\subsection{Special case: black-hole merger with one soft graviton}
\label{sec:SenMerger}

We now specialize the general result (\ref{eq:SenClassical}) to the case of a merger where only the massive states (i.e. the incoming black holes and resulting remnant) represent the hard sector. Let $p_1,p_2$ denote the incoming black-hole momenta, and let $p_X$ and $k$ denote the momenta of the outgoing remnant and graviton. Thus the all-incoming kinematics used later in the amplitude calculation is
\begin{equation}
 p_1+p_2+p_X+k=0,
 \qquad
 \omega_a:=p_a\cdot k,
 \qquad
 \omega_1+\omega_2+\omega_X=0,
 \label{eq:SenAllIncomingKinematics}
\end{equation}
with $\varsigma_1=\varsigma_2=+1$ and $\varsigma_X=-1$.  In these all-incoming variables, the physical Weinberg factor of the merger is
\begin{equation}
 \Stree{-1}{k}
 :=-\frac{\kappa}{2}\sum_{a=1,2,X}
 \frac{(p_a\cdot\epsilon_k^\sigma)^2}{\omega_a}.
 \label{eq:AllIncomingSoftRelation}
\end{equation}
The leading part of the classical soft theorem result therefore predicts
\begin{equation}
 S_{\rm gr}^{\rm merger}\Big|_{\omega^{-1}}
 =\Stree{-1}{k}.
 \label{eq:SenMergerWeinberg}
\end{equation}

At the first non-trivial logarithmic order the general formula simplifies considerably.  The propagation term contains only the outgoing massive remnant and becomes
\begin{equation}
 \left.S_{\rm gr}^{\rm merger}\right|_{\rm drag}
 =\frac{i\kappa^3}{32\pi}\,
 \frac{K_\sigma^2}{\omega_1\omega_2}
 \bigl(\ln\omega^{-1}+\ln R^{-1}\bigr)\,,
 \label{eq:SenMergerDrag}
\end{equation}
with
\begin{align}
    K_{\sigma}=\omega_2\,(p_1\!\cdot\!\epsilon_k^{\sigma})-\omega_1\,(p_2\!\cdot\!\epsilon_k^{\sigma})
    \,.
\end{align}
For the acceleration term, the only same-side massive pair is the incoming pair $(1,2)$.  Writing $y_{12}:=p_1\cdot p_2$ and using the specialization $D_{12}$ of Eq.~\eqref{eq:Dab}, we obtain
\begin{equation}
 \left.S_{\rm gr}^{\rm merger}\right|_{\rm acc}
 =\frac{i\kappa^3}{64\pi}\,
 \frac{K_\sigma^2}{\omega_1\omega_2}
 \frac{y_{12}(2y_{12}^2-3m_1^2m_2^2)}{D_{12}^{3}}
 \ln\omega^{-1}.
 \label{eq:SenMergerAcc}
\end{equation}
The $\ln R^{-1}$ piece in Eq.~\eqref{eq:SenMergerDrag} is the long-distance propagation phase associated with the location of the detector.  Thus the total $\ln\omega^{-1}$ prediction yields:
\begin{equation}
 \boxed{
 \left.S_{\rm gr}^{\rm merger}\right|_{\ln\omega^{-1}}
 =\frac{i\kappa^3}{64\pi}\frac{K_\sigma^2}{\omega_1\omega_2}
 \left[
 2\ln(\omega+i\varepsilon)^{-1}+\frac{y_{12}(2y_{12}^2-3m_1^2m_2^2)}{D_{12}^{3}}\ln(\omega-i\varepsilon)^{-1}
 \right]}
 \label{eq:SenNLOtarget}
\end{equation}
The two terms in the square bracket have distinct physical origins: the first is gravitational drag of the emitted graviton by the remnant, while the second is the radiation caused by the early-time acceleration of the two incoming black holes.  

\subsection{Perturbative organization of the soft expansion}
\label{sec:SenPredictions}

The classical soft theorem itself is an expansion in the emitted frequency $\omega$, for a generic physical process. For our purpose, however, it is useful to restore the powers of $\kappa=\sqrt{32\pi G}$ and then organize the expansion order by order in the gravitational interaction. This converts the low-frequency expansion into a bookkeeping double expansion. In the normalization of Eq.~\eqref{eq:SenSoftDefinition}, its schematic structure is
\begin{align}
 S_{\rm gr}^{\rm merger}(\omega,\kappa)
 ={}&\kappa\left[
 \frac{s^{(-1)}_{0}}{\omega}
 +s^{(0)}_0
 +\omega s^{(1)}_0+\cdots\right]
 \nonumber\\
 &+\kappa^3\left[
 \frac{s^{(-1)}_{1}}{\omega}
 +s^{(\log)}_{1}\ln\omega^{-1}
 +s^{(0)}_1+\cdots\right]
 +\mathcal O(\kappa^5).
 \label{eq:SenDoubleExpansion}
\end{align}
The meaning of the coefficients is as follows.
\begin{itemize}
 \item At order $\kappa/\omega$, the coefficient is the three-hard-leg Weinberg memory in Eq.~\eqref{eq:SenMergerWeinberg}.  At the same leading coupling, the ordinary tree-level soft expansion also supplies the familiar angular-momentum terms at orders $\omega^0$ and $\omega^1$~\cite{Laddha:2017ygw,LaddhaSen2018GravityWaves}.  

 \item At order $\kappa^3/\omega$, this stems from the previous discussion of the backreaction from the finite-energy gravitational radiation in the leading soft theorem.  The resulting coefficient contains recoil of the remnant together with the non-linear (Christodoulou) memory.  Its form is constrained by the leading soft theorem, but its numerical value depends on the radiated energy distribution.

 \item There could be a superclassical $\kappa^3\,\omega^{-1}\ln\omega$ term.  Such a term is more singular than the logarithmic soft theorem allows and must cancel in a consistent calculation~\cite{Laddha:2018vbn,Sahoo:2018lxl}.

 \item The coefficient $s^{(\log)}_{1}$ is universal and is completely fixed by the hard asymptotic data~\cite{Laddha:2018myi,Sahoo:2018lxl,Saha:2019tub,Krishna:2023fxg,Agrawal:2023zea,Boschetti:2026gfd}.  For the merger it is given explicitly by Eq.~\eqref{eq:SenNLOtarget}.  This is the sharpest prediction for the logarithmic part of the NLO waveform.

 \item Analytic terms of order $\kappa^3\omega^0$ and higher are not fixed by the logarithmic theorem and may depend on the short-distance merger dynamics. 
\item The non-analytic terms of the form $\omega^{n-1} (\ln \omega)^n$ have been derived explicitly for $n=2$ in the generic classical gravitational scattering, while their extension to $n\geq 3$ is presently conjectural in the general case~\cite{Sahoo:2020ryf,AtulBhatkar:2020hqz,Boschetti:2025tru,Banerjee:2026keq}. Other non-analytic pieces of the form $\omega^{n-1} (\ln \omega)^{r<n}$ are not universally fixed by the known classical soft theorem.
\end{itemize}

For the remainder of the paper, Eqs.~\eqref{eq:SenMergerWeinberg} and~\eqref{eq:SenNLOtarget}, together with the absence of a $\kappa^3\omega^{-1}\ln\omega$ term and the separate $\kappa^3/\omega$ non-linear-memory contribution, will be treated as the classical-soft-theorem benchmarks.  In the next section, we define radiation observables purely based on the amplitudes and reproduce all these features without any reference to classical physics.

\textbf{Comparison with previous approaches:} Finally, Refs.~\cite{Sahoo:2018lxl,Sen:2024qzb} also discuss the derivation of the classical soft theorem based on quantum scattering amplitudes. Although the classical waveform \eqref{eq:SenClassical} is obtained by both classical and quantum computations, they found additional logarithmic structures in the one-loop quantum soft factor which are not present in the classical result:
\begin{equation}
\begin{aligned}
\Delta S_{\rm gr}
&=\frac{\kappa^3}{128\pi^2}\ln \omega^{-1}
\sum_a \frac{\varepsilon_{\mu \rho}p_a^{\rho}k_{\nu}}{p_a \cdot k}
\left(
p_a^{\mu} \frac{\partial}{\partial p_{a,\nu}}
-p_a^{\nu} \frac{\partial}{\partial p_{a,\mu}}
\right)
\\[-1mm]
&\qquad\times
\sum_{a\neq b}
\frac{2(p_a\cdot p_b)^2-m_a^2m_b^2}{D_{ab}}
\ln \frac{p_a \cdot p_b + D_{ab}}
         {p_a \cdot p_b - D_{ab}}
\\
&\quad
+\frac{\kappa^3}{64\pi^2}
(\ln \omega^{-1}+\ln R^{-1})
\sum_a \frac{\varepsilon_{\mu\nu}p^{\mu}_a p^{\nu}_a}{p_a \cdot k}
\sum_b p_b \cdot k
\ln \frac{m_b^2}{(p_b\cdot \bar{k})^2}
\,.
\end{aligned}
\label{Sgr_quantum}
\end{equation}
According to the $\hbar$ counting, these additional terms can be comparable to the universal classical results \eqref{eq:SenClassical}. Thus, there would be no a priori reason to regard Eq.~\eqref{Sgr_quantum} as ``quantum''. Understanding the absence of this contribution constitutes another objective of the present work. Indeed, any consistent microscopic description of black hole mergers must explain why Eq.~\eqref{Sgr_quantum} does not contribute to the classical waveform.

\section{Inclusive merger waveforms from KMOC and soft theorems}
\label{sec:KMOC}

We consider an initial condition in which two Schwarzschild black holes are initially far separated and will collide to form a Kerr black hole whilst emitting gravitational waves. In the KMOC formalism~\cite{Kosower:2018adc,Cristofoli:2021vyo}, the initial state is given by two wavepacket states 
\begin{align}
\ket{\Psi}:=\int\dd\Phi(p_1)\,\dd\Phi(p_2)\,\phi_1(p_1)\phi_2(p_2)\,e^{i(b_1\cdot p_1+b_2\cdot p_2)}\ket{p_1;p_2}\,,
\label{eq:Psi}
\end{align}
where $p_a~(a=1,2)$ are spin-0 particles representing Schwarzschild black holes and $b=b_1-b_2$ is the impact parameter. The wavefunction $\phi_a$ is assumed to be sharply peaked at the classical momentum $m_a u_a$ with $u^2=1$ and to satisfy the so-called Goldilocks relation to have a well-defined classical limit~\cite{Kosower:2018adc}. The outgoing Hilbert space decomposes into hard sectors tensored with the graviton Fock space, 
\begin{equation}
\mathcal H_{\rm out}=\bigoplus_A\left(\mathcal H_A^{\rm hard}\otimes\mathcal F_{\rm grav}\right),
\qquad
\mathcal F_{\rm grav}=\bigoplus_{n=0}^{\infty}\mathcal H_n^{\rm rad}\,.
\label{eq:HilbertDecomp}
\end{equation}
Here $A$ labels the hard final-state sector, while $n$ labels the graviton number within the radiation Fock space. In the classical limit, the relevant hard sectors are either two-massive-particle states (scattering, $A=1,2$) or one-massive-particle states (merger, $A=X$), the latter of which we are interested in. In this section, we elaborate on how this problem is described by scattering amplitudes and how emitted gravitational waves are computed.

\subsection{Effective merger amplitudes in the coherent-spin basis}
\label{sec:matching}

The on-shell amplitude for merging two spin-0 particles into a spin-$j$ remnant $X$ of mass $m_X$ is entirely fixed by little-group covariance up to a constant,
\begin{align}
\Amp^{I_1\cdots I_{2j}}(p_X,j|p_1;p_2)=g_j\,\bra{\bm X}p_1p_2\ket{\bm X}^{j}\,,
\label{3pt_j}
\end{align}
in the massive spinor-helicity formalism~\cite{Arkani-Hamed:2017jhn}. The remnant is a continuum state, described by a spectral density $\rho_j(m_X^2)$. The combination $\rho_j|g_j|^2$ is fixed by imposing that the $X$-production channels saturate unitarity, i.e., the black-disk absorption~\cite{Aoki:2024boe}.

Let us first recap the case where graviton emission is neglected~\cite{Aoki:2024boe}. The partial absorption cross section for the $j$-th partial wave for the single $X$ state is
\begin{align}
   \sigma^{(0)}_j&=\frac{1}{4EP}\int\dd m_X^2\,\rho_j(m_X^2)\int\dd\Phi(p_X)\,\hat{\delta}^{(4)}(p_{12}+p_X)
\sum_{I_1\cdots I_{2j}}\bigl|\Amp^{I_1\cdots I_{2j}}(p_X,j|p_1;p_2)\bigr|^2\,,
\label{eq:sigmaj}
\end{align}
while the black-disk cross-section is
\begin{align}
\sigma^{\rm BD}_j=\frac{\pi(2j+1)}{P^2}\,\theta(1-j/L_c)\,.
\label{eq:blackdiskj}
\end{align}
Here, $E$ and $P$ are the centre-of-mass energy and momentum
\begin{align}
    E=\sqrt{p_{12}^2}\,, \quad P=\sqrt{\frac{(p_1 \cdot p_2)^2-m_1^2m_2^2}{p_{12}^2}} = \sqrt{ \frac{\lambda(p_{12}^2,m_1^2, m_2^2)}{4p_{12}^2} } \,,
\end{align}
where $p_{12}:=p_1+p_2$ and $\lambda(x,y,z):=x^2+y^2+z^2-2xy-2yz-2zx$ is the K\"{a}ll\'{e}n function. $L_c$ is the critical angular momentum for whether two particles collide. In the case of black hole collision, the critical value is $L_c=b_cP\sim Gm_1m_2$. Then, we impose the condition that $\sigma^{(0)}_j=\sigma^{\rm BD}_j$ for $j<L_c$ which fixes $\rho_j|g_j|^2$. We adopt the convention\footnote{Physically, the spectral density represents internal degrees of freedom of black holes and the couplings are the ones for microstates. While this distinction is important when discussing microscopic properties~\cite{Akpinar:2026oni}, they appear in the combination $\rho_j |g_j|^2$ in classical computations. Therefore, we choose the assignment of $\rho_j$ and $g_j$ as a convention for convenience, and we specifically choose it to exponentiate the three-point amplitude.}
\begin{equation}
g_j=m_X\frac{\sqrt{(2j)!}}{j!}\bigl[m_X^2\,\lambda(m_1^2,m_2^2,m_X^2)\bigr]^{-j/2}\,,
\label{eq:gj}
\end{equation}
for which the tensor-product structure of \eqref{3pt_j} resums into an exponential in the coherent-spin basis~\cite{Aoude:2021oqj}, labelled by the SU(2) variables $(\alpha^I,\tilde\alpha_I)$,
\begin{align}
&\Amp(p_X,\tilde\alpha|p_1;p_2)=m_X\,e^{-\frac12\|\alpha\|^2+z}\,,\qquad
z:=\frac{\tilde\alpha_I\bra{p_X^I}p_1p_2\ket{p_X^J}\tilde\alpha_J}{m_X\,\lambda^{1/2}(m_1^2,m_2^2,m_X^2)}\,,
\label{3pt_alpha}
\\
&\rho_{\alpha}= \frac{4 \|\alpha \|^2 }{\lambda^{1/2}(m_1^2,m_2^2, m_X^2)}\, \theta(1-\|\alpha\|^2/a_c)
\,, \qquad a_c:=2L_c
\,.
\label{rhoalpha}
\end{align}
The coherent-spin state $\bra{p_X,\alpha}$ is a superposition of spin-$j$ states. The Pauli-Lubanski pseudovector for $\bra{p_X, \alpha}$ is given by
\begin{align}
S_X^{\mu}= \bra{p_X, \tilde{\alpha}} \hat{S}^{\mu}_X \ket{p_X, \alpha}  =  \frac{\hbar}{2} \tilde{\alpha}_I [\sigma^{\mu}_X]^I{}_J \alpha^J
\,,
\label{SX_def}
\end{align}
with
\begin{align}
[\sigma^{\mu}_X]^I{}_J := \frac{1}{2m_X}\left( \bra{X^I}\sigma^{\mu}|X_J] + [X^I|\bar{\sigma}^{\mu}\ket{X_J} \right)
\,.
\end{align}
Eq.~\eqref{SX_def} is identified with the spin vector of the formed Kerr black hole in the classical limit. The exponentiated structure of \eqref{3pt_alpha} is crucial to correctly reproduce the classical angular momentum conservation for black hole mergers explained below.

\subsection{Inclusive radiative matching}
\label{subsec:inclusive_radiative_matching}

When graviton emissions are taken into account, the combination $\rho_j |g_j|^2$ receives associated radiative corrections. The matching condition is, diagrammatically,
\begin{equation}
\;\Biggl|\;\;\vcenter{\hbox{\begin{tikzpicture}[cutdiag]
\coordinate (V) at (0,0);
\extleg[sc]{V}{-0.60,0.42}{$1$}
\extleg[sc]{V}{-0.60,-0.42}{$2$}
\extleg[ms]{V}{0.78,0}{$X$}
\end{tikzpicture}}}\;\;\Biggr|^{2}
+\;\Biggl|\;\;\vcenter{\hbox{\begin{tikzpicture}[cutdiag]
\coordinate (V) at (0,0);
\extleg[sc]{V}{-0.62,0.42}{$1$}
\extleg[sc]{V}{-0.62,-0.42}{$2$}
\extleg[gr]{V}{0.74,0.40}{$\ell,\,{\eta}$}
\extleg[ms]{V}{0.74,-0.34}{$X$}
\node[blob, minimum size=4.6mm] at (V) {};
\end{tikzpicture}}}\;\;\Biggr|^{2}
+\cdots
\;=\;\sigma^{\rm BD}_j\,.
\label{eq:sigmabdiag}
\end{equation}
That is, unitarity is saturated by the total transition probability to $\mathcal{H}^{\rm hard}_X\otimes \mathcal{F}_{\rm grav}$. In practice, however, it is cumbersome to impose the condition for each partial wave $j$, as we need to resum all spinning states to move to the coherent-spin basis. We thus reformulate it into a way to directly impose conditions for coherent-spin amplitudes, rather than discrete spins.

For convenience, we define the transition amplitude for initial plane-wave states and wavepacket states:
\begin{align}
i\Gamma^{\eta_1\cdots\eta_n}_{3+n}(\ell_1,\dots,\ell_n)
&:=\bra{p_X,\alpha;\ell_1^{\eta_1};\cdots;\ell_n^{\eta_n}}S\ket{p_1;p_2}\,,
\nn
i\widetilde\Gamma^{\eta_1\cdots\eta_n}_{3+n}(\ell_1,\dots,\ell_n)
&:=\bra{p_X,\alpha;\ell_1^{\eta_1};\cdots;\ell_n^{\eta_n}}S\ket{\Psi}\,,
\label{eq:Gammadef}
\end{align}
with $\ell_r,\eta_r$ the momenta and helicities of the emitted gravitons; stripping the momentum-conserving delta function gives the amplitudes, $\Gamma_3=\Amp(p_X,\alpha|p_1;p_2) \hdelta^{(4)}(p_1+p_2+p_X)$. The $(3+n)$-point amplitude starts at $\mathcal{O}(\kappa^n)$. In terms of the smeared amplitudes with definite impact parameter, our condition is simply that the total probability to $\mathcal{H}^{\rm hard}_X\otimes \mathcal{F}_{\rm grav}$ is unity below the critical impact parameter\footnote{Once radiative effects are included, the critical impact parameter $b_c$ itself should also receive radiative corrections, since the emitted radiation modifies the energy and angular momentum available to the hard merger process.}:
\begin{align}
    P_{\rm abs}= \theta(b_c-b)
    \label{P_matching}
\end{align}
where
\begin{align}
P_{\rm abs}&:=\sum_{n=0}^{\infty}P^{(n)}_{\rm abs}\,,\qquad
P_{\rm abs}^{(n)}:=\frac{1}{n!}\sum_{\eta_1\cdots\eta_n}\int_{X,\ell^n}\bigl|\widetilde\Gamma^{\eta_1\cdots\eta_n}_{3+n}\bigr|^2\,,
\label{eq:P}
\end{align}
with the shorthand for the on-shell measures
\begin{align}
\int_X:=\int\frac{\dd^2\alpha\,\dd^2\tilde\alpha}{\pi^2}\,\dd m_X^2\,\dd\Phi(p_X)\,\rho_\alpha(m_X^2)\,,\qquad
\int_\ell:=\int\dd\Phi(\ell)\,,\qquad
\int_{X,\ell^n}:=\int_X\prod_{r=1}^{n}\int_{\ell_r}\,.
\label{Xmeasure}
\end{align}

Before adding radiation, let us recall how the classical limit of \eqref{3pt_alpha} arises, as it will be repeatedly used below. Reintroducing $\hbar$ with the scalings $\alpha \sim \tilde{\alpha}\sim \hbar^{-1/2}$ and $\Amp_3\to\frac{1}{\hbar^{1/2}}\Amp_3$, the real part of the superclassical $\mathcal{O}(\hbar^{-1})$ piece of the exponent must vanish for the amplitude to survive the $\hbar\to0$ limit, turning the three-point into Gaussian representations of delta functions (see Appendix~\ref{sec:hbarApp} for a review of $\hbar$ counting). Concretely, in the centre-of-mass frame with the spatial coordinates
\begin{align}
p_1^{\mu}&=(E_1, P, 0, 0) \,,  \quad
p_2^{\mu}=(E_2,-P,0, 0) \,,  \nn
b_1^{\mu}&=\left(0, 0, -\frac{E_2}{E}b, 0 \right)\,, \quad
b_2^{\mu}=\left(0,0, \frac{E_1}{E}b, 0 \right) \,, \label{kinematics} \\
b^{\mu}&=b_1^{\mu}-b_2^{\mu}=(0, 0, -b, 0)
\,, \nonumber
\end{align}
one finds~\cite{Aoki:2024boe}
\begin{align}
\Amp = \frac{m_X}{\hbar^{1/2}}\, e^{-\frac{1}{2{\hbar}}\|\alpha \|^2 +z}  = \hbar^{1/2} \pi E\, \delta_{\hbar} (\ima \alpha^1 )\,\delta_{\hbar} (\re \alpha^2)\, e^{\mathcal{O}(\hbar^0)}
\,, \qquad
\delta_{\hbar}(x) := \frac{e^{-x^2/\hbar}}{\sqrt{\pi \hbar}} \to \delta(x)
\,.
\label{3pt_delta}
\end{align}
On the support of these delta functions, the spin of the $X$ state is transverse to the momenta $p_{1}$ and $p_2$,
\begin{align}
S^{\mu}_X=\left( 0, 0, (\re \alpha^1)(\ima \alpha^2), \frac{1}{2}[(\ima \alpha^2)^2 - (\re \alpha^1)^2] \right)
\,.
\label{comSX}
\end{align}
The remaining components are fixed by the Fourier transform to impact-parameter space, appearing in the wavepacket state in \eqref{eq:Psi}, by keeping the exponent at $\mathcal{O}(\hbar^0)$ in \eqref{3pt_delta}. Then, in the classical limit, \eqref{3pt_alpha} and \eqref{rhoalpha} yield~\cite{Aoki:2024boe}
\begin{align}
    P^{(0)}_{\rm abs}&=\int_X |\widetilde{\Gamma}_3|^2
    \nn
    &=\frac{1}{2}\int\dd^2 \alpha \dd^2 \tilde{\alpha} \|\alpha \|^2 \delta_{\hbar} (\ima \alpha^1 )\,\delta_{\hbar} (\re \alpha^2)\,   \delta(S^z_X - b P)\, \delta(S^y_X)=1
    \,,
\end{align}
for $b=|S_X|/P<b_c$. In short, thanks to the exponential of \eqref{3pt_alpha}, the combination $\rho_{\alpha}|\widetilde{\Gamma}_3|^2$ localises $\alpha$ such that the spin vector $S_X$ satisfies the classical angular momentum conservation $\sum_a J_a^{\mu\nu}=0$. This can be generalised to computing expectation values as long as they do not change the exponential structure for the classical localisation
\begin{align}
    \int_X O(p_a,S_X) |\widetilde{\Gamma}_3|^2 = O\Big|_{\sum_a p_a=0,~\sum_a J_a=0,~p_a=m_a u_a}
    \,.
    \label{eq:classical_localization}
\end{align}
This simplicity reflects how the conservation law entirely determines three-point classical dynamics.

In the presence of gravitational radiation, the exclusive merger channel
$1+2\to X$ no longer saturates the absorption probability by itself. In
addition to virtual gravitational corrections to the $1+2\to X$ transition,
real-emission channels
\begin{equation}
    1+2\to X +g \,,
    \qquad
    1+2\to X + g + g\,,
    \qquad \ldots
\end{equation}
open perturbatively in $\kappa$. Correspondingly, the normalization of the
exclusive zero-graviton merger amplitude receives radiative corrections.
We parameterize the part of this radiative dressing that preserves the
coherent-spin structure of the leading merger amplitude by
\begin{align}
    \Amp_3
    &=
    Z(z)\,m_X\,e^{-\frac12\|\alpha\|^2+z}\,,
    \qquad
    Z(z)=1+\mathcal{O}(\kappa^2)\,.
    \label{eq:Zdef}
\end{align}
Here $z$, defined in \eqref{3pt_alpha}, is a function of the coherent-spin
variable $\tilde{\alpha}$. The factor $Z$ should be regarded as the
radiative normalization of the exclusive $1+2\to X$ merger channel. We do not require
its full perturbative form below, we only assume that, to the order of interest, the
radiative dressing does not modify the exponential structure responsible
for the classical localization of the remnant spin.

On the support of the classical angular-momentum-conservation conditions,
the zero-graviton contribution to the absorption probability is therefore
\begin{align}
    P^{(0)}_{\rm abs}
    &=
    |Z|^2\,,
\end{align}
where $\tilde{\alpha}$ is evaluated on the classical support. For
$b<b_c$, perfect absorption requires the complete set of merger states
$X+\text{gravitons}$ to saturate unitarity, and hence
\begin{align}
    |Z|^2+\sum_{n\geq1}P^{(n)}_{\rm abs}
    &=
    1\,.
    \label{eq:inclusiveZ}
\end{align}
This relation is an inclusive real--virtual matching condition for the
merger.\footnote{This structure is analogous to the real--virtual probability balance underlying Sudakov or no-emission factors in gauge theory: virtual corrections reduce the probability for the exclusive zero-radiation channel, while the missing probability is carried by states with real radiation. We emphasize, however, that we do not assume a Sudakov exponentiation of $Z$; here the analogy refers only to the order-by-order unitarity relation between exclusive and radiative channels.} The reduction of the exclusive $1+2\to X$ probability due to the
radiative dressing of $Z$ is compensated by the probability carried by
states containing real gravitons.

To make this relation explicit, let
\begin{align}
    Z
    &=
    1+\kappa^2 z_2+\mathcal{O}(\kappa^4)\,.
    \label{eq:Zexpansion}
\end{align}
Since the one-graviton transition amplitude starts at
$\mathcal{O}(\kappa)$, its probability starts at
$\mathcal{O}(\kappa^2)$, whereas the two-graviton probability starts at
$\mathcal{O}(\kappa^4)$. Expanding \eqref{eq:inclusiveZ} therefore gives
\begin{align}
    |Z|^2-1+P^{(1)}_{\rm abs}
    &=
    \mathcal{O}(\kappa^4)\,,
    \label{eq:matchingZ}
\end{align}
or, equivalently,
\begin{align}
    2\,\mathrm{Re}\,z_2
    &=
    -\left.
    \frac{P^{(1)}_{\rm abs}}{\kappa^2}
    \right|_{\mathcal{O}(\kappa^0)}\,.
    \label{eq:Zunitarity}
\end{align}
Thus, at this order, the correction to the modulus of the exclusive merger
amplitude does not constitute independent merger data: it is fixed by the
inclusive one-graviton emission probability.

This matching relation will be particularly useful for the radiation
observable below. At the order considered here, the KMOC waveform contains
the leading soft factor multiplied by the inclusive combination
$|Z|^2+P^{(1)}_{\rm abs}$. Equation~\eqref{eq:matchingZ} therefore removes
the unknown radiative normalization of the exclusive merger amplitude from
the NLO waveform. What survives are nontrivial inclusive moments of the
radiation, such as the radiated momentum and energy distribution, which
give rise to recoil and non-linear memory.

\subsection{Radiation observables from KMOC}
\label{sec:KMOCwaveform}
In the KMOC formalism, the gravitational wave is an expectation value of the graviton operator at future null infinity~\cite{Kosower:2018adc,Cristofoli:2021vyo}:
\begin{equation}
h_{\mu\nu}(x)=\sum_\sigma\kappa\int\dd\Phi(k)\,e^{-ik\cdot x}\,\varepsilon^{-\sigma}_{k,\mu\nu}\,i\W_\sigma(k)+\text{c.c.},
\label{hmunu}
\end{equation}
where the spectral waveform is
\begin{equation}
i\W_\sigma(k)=\bra{\Psi}\hat{S}^\dagger \hat{a}_\sigma(k)\hat{S}\ket{\Psi}\,.
\label{eq:KMOCkerneldef}
\end{equation}
Note that the KMOC waveform $\mathcal{W}$ and the classical-soft-theorem waveform $S_{\rm gr}$ are related by $i\mathcal{W}=S_{\rm gr}$. Inserting the complete set of merger states $\mathcal{H}_X\otimes \mathcal{F}_{\rm grav}$, the waveform is organised by the number of cut gravitons (i.e. states in $\mathcal{F}_{\rm grav}$),
\begin{align}
i\W_\sigma=\sum_{n=0}^{\infty} i\W^{(n)}_\sigma=i\mathcal{W}^{(0)}_{\sigma} +i\mathcal{W}^{(1)}_{\sigma} + (n\geq 2)\,,
\label{eq:Wn}
\end{align}
where 
\begin{align}
i\W^{(0)}_\sigma(k)
= \int_X \widetilde\Gamma_3^*\,\widetilde\Gamma^\sigma_4(k)\,,
\qquad
i\W^{(1)}_\sigma(k)
= \sum_\eta \int_{X,\ell}
\bigl[\widetilde\Gamma^\eta_4(\ell)\bigr]^*
\widetilde\Gamma^{\sigma,\eta}_5(k,\ell)\,.
\label{eq:W0W1}
\end{align}
The set of momenta of the graviton being integrated is denoted by $\ell$, and is always distinguished from the observed graviton momentum $k$. 
In addition, the momentum radiated by gravitational waves is computed by 
\begin{align}
    R^{\mu}=\int_\ell \ell^{\mu} \mathcal{N}(\ell)\,, \qquad \mathcal{N}=\sum_{\eta} \bra{\Psi}\hat{S}^{\dagger}\hat{a}^{\dagger}_{\eta}(\ell)\hat{a}_{\eta}(\ell)\hat{S}\ket{\Psi}
    \,,
    \label{eq:radiated_momentum}
\end{align}
where the number density of gravitons is
\begin{align}
    \mathcal{N}=\sum_{n \geq 1}\mathcal{N}^{(n)}=\sum_{\eta} \int_X |\widetilde{\Gamma}^{\eta}_4(\ell)|^2 + (n \geq 2)
    \,.
    \label{eq:graviton_number_density}
\end{align}

In the following sections, we use the soft theorems to reorganise the KMOC
waveform~\eqref{eq:KMOCkerneldef} and clarify the origin of gravitational
logarithmic tail, recoil, and non-linear memory in black-hole mergers.\footnote{
For related applications of soft theorems to classical observables, see e.g., Refs.~\cite{Bautista:2021llr,DiVecchia:2021ndb,Alessio:2024onn,Alessio:2024wmz,Paul:2026hyf,Blas:2026yqb}.
} It is therefore useful to first clarify
the hierarchy of scales and the perturbative organisation that we employ
below. The characteristic frequency of the merger is set by the inverse
size of the final black hole and is parametrically of the order of the ringdown
frequency,
\begin{align}
    \omega_*\sim \frac{1}{GM}\,.
\end{align}
We refer to the observed radiation as soft when its frequency is much
smaller than this scale, while radiation with frequency comparable to the
merger scale will be referred to as finite-frequency or hard radiation:
\begin{align}
    \omega_{\rm soft}\ll\omega_*\,,
    \qquad
    \omega_{\rm hard}\sim\omega_*\,.
    \label{eq:soft-hard-scales}
\end{align}
As the KMOC waveform is
an inclusive in-in observable, the momenta of the radiative gravitons
crossing the cut are integrated over their full physical range. Thus, even
when the observed momentum $k$ is soft, a cut graviton with momentum
$\ell$ may lie either in the double-soft region,
\begin{align}
    \ell\sim k\ll \omega_*\,,
\end{align}
or in the hierarchical region,
\begin{align}
    k\ll \ell\lesssim\omega_*\,.
\end{align}
In the former region both gravitons are governed by the double-soft
theorem, while in the latter only the observed graviton is soft and the
radiation should be regarded as part of the hard final
state.

Therefore, we employ the double expansions in the frequency of the observed graviton and in the number of unobserved gravitons of finite frequencies. This double expansion scheme can be connected to the universal $\kappa$-expansion\footnote{The gravitational constant $G$ is often used for bookkeeping of the perturbative expansion. They are related by $\mathcal{O}(G^n)=\mathcal{O}(\kappa^{2n-1})$ where the former counts the order of $G$ appearing in the dimensionless gravitational field $h_{\mu\nu}$ while the latter counts the order of $\kappa$ of the spectral waveform $\mathcal{W}_{\sigma}$.} as follows. Since the natural dimensionless combination is $GM\omega$, the low-frequency expansion combines the $\kappa$-expansion as $\omega_{\rm soft}=\mathcal{O}(\kappa^2)$. On the other hand, the hard graviton appears only from the integrals. One hard graviton integral requires two graviton emissions in total, one from $\bra{\Psi}\hat{S}^{\dagger}$ and another from $\hat{S}\ket{\Psi}$, yielding at least $\kappa^2$:
\begin{align}
    \int_{\ell_{\rm hard}} = \mathcal{O}(\kappa^2)
    \,.
\end{align}
Hence, when we refer to the waveform at $\mathcal{O}(\kappa^3)$, this is the one containing up to the subleading order in the soft expansion and the leading order in the backreaction of radiation.

\subsection{Soft theorems for wavepacket states}
\label{sec:classical_soft_thm}
As discussed in the previous section, the low-frequency expansion of the waveform is mapped, via the KMOC formalism, to the soft-momentum expansion of the gravitational $S$-matrix. Analogously to classical soft theorems, symmetries constrain this expansion universally up to a finite order in the soft momenta, with the extent of this universality dictated by the underlying symmetry. We will refer to these as \textit{quantum soft theorems}. For instance, we consider the four-point matrix element $\Gamma_4$ with one soft graviton. The soft graviton theorem reads~\cite{Weinberg:1965nx,Bern:2014oka}
\begin{align}
\Gamma_4^\sigma(k)
&=\hat S_k\,\Gamma_3
\,, \label{Gamma4_soft}
\\
\hat S_{k}
&=\left[\hStree{-1}{k} + \hStree{0}{k} +\cdots \right]
+\hSloop{0}{k}+\cdots\,.
\label{eq:softop}
\end{align}
where $\hStree{-1}{k}$ is the leading Weinberg soft factor \eqref{eq:AllIncomingSoftRelation} and $\hStree{0}{k}$ is the sub-leading soft factor
\begin{align}
    \hStree{0}{k} = \frac{i\kappa}{2} \sum_a\,\frac{(\epsilon^{\sigma}_k\!\cdot p_a)(\epsilon^{\sigma}_{k,\mu}\hat{\mathcal J}_a^{\mu\nu}k_\nu)}{p_a\cdot k}
\end{align}
with $\hat{\mathcal J}_a^{\mu\nu}$ being the Lorentz generator of the leg $a$ acting on $\Gamma_3=\Amp_3 \hdelta^{(3)}(p_1+p_2+p_X)$. $\hSloop{0}{k}$ denotes the one-loop correction to the soft factor~\cite{Sahoo:2018lxl}. Eq.~\eqref{Gamma4_soft} should be understood in the sense of distributions, as it relates the four-point kinematics $p_1+p_2+p_X+k=0$ to the three-point kinematics $p_1+p_2+p_X=0$ under the soft limit. The soft factor $\hat{S}_k$ is an operator, acting on the delta function as well as the amplitude. On the other hand, in the classical limit, the quantum operator $\hat{\mathcal{J}}^{\mu\nu}_a$ can be expected to be replaced with the classical angular momentum $\mathcal{J}^{\mu\nu}_a$ of the associated hard particle. In this section, we discuss how this prescription naturally arises by considering the wavepacket amplitude $\widetilde{\Gamma}_{3+n}$.

The incoming particles 1 and 2 are spinless, so that the Lorentz generator of $a=1,2$ is purely orbital
\begin{align}
\hat{\mathcal{J}}^{\mu\nu}_{1,2}=\hat L^{\mu\nu}_{1,2}\,,\qquad
\hat L^{\mu\nu}_a:=i\left(p_a^\mu\frac{\partial}{\partial p_{a,\nu}}-p_a^\nu\frac{\partial}{\partial p_{a,\mu}}\right)\,.
\label{eq:Lop}
\end{align}
The Lorentz generator of $X$ contains both orbital and spin parts. Lorentz invariance $\sum_a \hat{\mathcal{J}}_a^{\mu\nu}\Gamma_3=0$ allows us to write $\hat{\mathcal{J}}_X^{\mu\nu}$ in terms of $\hat{L}^{\mu\nu}_{1,2}$:
\begin{align}
\hat{\mathcal{J}}^{\mu\nu}_X\,\Gamma_3=-\left(\hat L^{\mu\nu}_1+\hat L^{\mu\nu}_2\right)\Gamma_3\,.
\label{eq:WardJ}
\end{align}
We then consider the wavepacket amplitude
\begin{align}
    \widetilde{\Gamma}_4(k)&=\int_{p_1,p_2}\phi_1(p_1)\phi_2(p_2)e^{i(b_1\cdot p_1+ b_2\cdot p_2)}\Gamma_4^{\sigma}(k)
    \nn
    &=\int_{p_1,p_2}\phi_1(p_1)\phi_2(p_2)e^{i(b_1\cdot p_1+ b_2\cdot p_2)} \hat{S}_k \Gamma_3
    \,.
\end{align}
Note that the wavepacket amplitude is an ordinary function, rather than a distribution, thanks to the Fourier integrals. Performing integration by parts, it is possible to arrange the operators so that they act on $\prod_{a=1,2}\phi_a e^{i b_a \cdot p_a}$ rather than the amplitude $\Gamma_3$. Under the Goldilocks relation, the variation of the wavefunction $\phi_a$ is smoother than $e^{i b_a \cdot p_a}$. Therefore, in the classical limit, we only need derivatives acting on $e^{i b_a \cdot p_a}$; concretely, for the subleading soft graviton operator, we have
\begin{align}
    &\int_{p_1,p_2}\phi_1(p_1)\phi_2(p_2)e^{i(b_1\cdot p_1+ b_2\cdot p_2)} i \frac{(\epsilon^{\sigma}_k\!\cdot p_a)(\epsilon^{\sigma}_{k,\mu}\hat{\mathcal J}_a^{\mu\nu}k_\nu)}{p_a\cdot k} \Gamma_3 
    \nn
    &= \int_{p_1,p_2}\phi_1(p_1)\phi_2(p_2) i\Gamma_3 \frac{(\epsilon^{\sigma}_k\!\cdot p_a)(-\epsilon^{\sigma}_{k,\mu}\hat{\mathcal J}_a^{\mu\nu}k_\nu)}{p_a\cdot k} e^{i(b_1\cdot p_1+ b_2\cdot p_2)}
    \nn
    &=\int_{p_1,p_2}\phi_1(p_1)\phi_2(p_2) i\Gamma_3 \frac{(\epsilon^{\sigma}_k\!\cdot p_a)(\epsilon^{\sigma}_{k,\mu}\mathcal J_a^{\mu\nu}k_\nu)}{p_a\cdot k} e^{i(b_1\cdot p_1+ b_2\cdot p_2)}
\end{align}
where
\begin{align}
   \mathcal J_1^{\mu\nu}=L_1^{\mu\nu}\,,\quad
\mathcal J_2^{\mu\nu}=L_2^{\mu\nu}\,,\quad
\mathcal J_X^{\mu\nu}=L_1^{\mu\nu}+L_2^{\mu\nu}\,,
\end{align}
and
\begin{align}
    L_a^{\mu\nu}=b_a^\mu p_a^\nu-b_a^\nu p_a^\mu
    \,.
\end{align}
Furthermore, since the wavefunctions are sharply peaked at four-momenta $p_a=m_a u_a$, we move the soft factor to the outside of the integral by evaluating it at $p_a=m_a u_a$. As a result, in the classical limit, the soft theorem for the wavepacket amplitude is written as
\begin{align}
\widetilde\Gamma_4^\sigma(k)
  &= S_k\,\widetilde\Gamma_3\,,
\qquad
S_k
  = \hat S_k\Big|_{\hat{\mathcal J}_a\to\mathcal J_a, \, p_a\to m_a u_a}
\label{eq:Scl}
\end{align}
where $S_k$, which is a c-number rather than an operator, is precisely the classical version of the soft factor. We will refer to this version of the soft theorem as \textit{wavepacket soft theorems} in the following analysis. 

In addition, for computing the waveform at $\mathcal{O}(\kappa^3)$, we need the double soft graviton theorem~\cite{Cachazo:2015ksa,Saha:2016kjr,Saha:2017yqi,Chakrabarti:2017ltl}:
\begin{align}
\Gamma_5^{\sigma,\eta}(k,\ell)=\hat S_{k,\ell}\,\Gamma_3
\quad \text{for} \quad k\sim \ell \ll p_a.
\label{eq:5pt_soft}
\end{align}
The universal part of the double soft factor can be written as\footnote{The Compton terms $\sum \Amp^{\text{c}}/p_a\cdot (k+\ell)$ agree with the gauge-invariant contact terms denoted by $\mathcal{A}_2$ in \cite{Chakrabarti:2017ltl} on the support of the momentum conservation of the hard amplitude $\Gamma_3$. }
\begin{align}
    \hat{S}_{k,\ell}&=\hStree{-2}{k,\ell} + \hStree{-1}{k,\ell}
    \,,     \label{eq:doublesoft} \\
    \hStree{-2}{k,\ell}&=\hStree{-1}{k} \hStree{-1}{\ell}
    \,, \\
    \hStree{-1}{k,\ell}&=
    \left(\hStree{-1}{k} \hStree{0}{\ell}  + \frac{\kappa}{2}\frac{(\epsilon^{\sigma}_k\cdot \ell)^2}{k\cdot \ell}\hStree{-1}{\ell} +(k\leftrightarrow \ell)\right) + \sum_a \frac{\Amp^{\text{c}}(p_a;k^{\sigma},\ell^{\eta})}{p_a\cdot (k+\ell)} 
    \,,
\end{align}
where $\Amp^{\text{c}}$ is the gravitational Compton amplitude~\cite{Arkani-Hamed:2017jhn},
\begin{align}
\Amp^{\text{c}}(p_a;k^+,\ell^+)
&=\left(\frac{\kappa}{2}\right)^2\frac{m_a^4\,[k\ell]^4}{(2\ell\cdot k)(2p_a\cdot k)(2p_a\cdot\ell)}\,, \nn
\Amp^{\text{c}}(p_a;k^-,\ell^{+})
&=\left(\frac{\kappa}{2}\right)^2\frac{\langle k|p_a|\ell]^4}{(2 \ell\cdot k)(2 p_a\cdot k)(2 p_a\cdot \ell)}\,.
\label{eq:contacthel}
\end{align}
See Appendix~\ref{app:BCFW} for an on-shell derivation of the spinless part of the double soft factor using BCFW construction. We then denote its wavepacket version by
\begin{align}
\widetilde\Gamma_5^{\sigma,\eta}(k,\ell)=S_{k,\ell}\,\widetilde\Gamma_3\,,~~
S_{k,\ell}=\hat S_{k,\ell}\Big|_{\hat{\mathcal J}_a\to\mathcal J_a,\, p_a \to m_a u_a}\,.
\label{eq:Gamma5cl}
\end{align}

\subsection{Waveforms from soft theorems and KMOC formalism}
\label{sec:memoryKMOC}
We now assemble the waveform at $\mathcal{O}(\kappa^3)$ from the wavepacket soft theorem. The $n=0$  piece is simply given by,
\begin{align}
    i\mathcal{W}^{(0)}_{\sigma}=\int_X S_k\,|\widetilde\Gamma_3|^2=|Z|^2 S_k
    \,.
\end{align}
As we have explained in Sec.~\ref{sec:KMOCwaveform}, it is crucial to split the $\ell$-integral into the double-soft and hierarchical-soft regions and separately deal with these integrals. In particular, in the hierarchical-soft region, the $\ell$ dependence is not governed by soft theorem as $\ell$ is integrated over finite frequencies. We split the integral by means of the method of regions~\cite{Beneke:1997zp, Smirnov:2002pj}:
\begin{align}
\int_\ell=\int_{\rm double}+\int_{\rm hierarchical}\,,\quad
{\rm double}:\ \ell\sim k \ll p_a\,,~~
{\rm hierarchical}:\ k\ll \ell \ll p_a\,.
\label{eq:softregions}
\end{align}
Note that, in the method of regions, while the integrand is expanded according to the momentum scaling of each region, the integral itself is integrated over the entire domain with dimensional regularisation. 

In the double soft region, the double soft graviton theorem is applied. On the other hand, one must use the single soft graviton theorem in a sequential order in the hierarchical region. Hence, the $(n=1)$-cut graviton contribution of the waveform is
\begin{align}
    i\mathcal{W}^{(1)}_{\sigma}&=\underbrace{\sum_{\eta}\int_{X,\ell}\bigl[S_{\ell} \widetilde\Gamma_3\bigr]^*
S_{k,\ell}\widetilde\Gamma_3}_{\rm double}
+\underbrace{\sum_{\eta}\int_{X,\ell}\bigl[\widetilde\Gamma^\eta_4(\ell)\bigr]^*
S_k(p_a,\ell) \widetilde\Gamma^{\eta}_4(\ell)}_{\rm hierarchical}
\,.
\label{eq:Wsplit}
\end{align}
The double-soft region can be simply reduced to,
\begin{equation}
i \mathcal{W}^{(1)}_{\sigma}|_{\rm double}=|Z|^2 \sum_{\eta} \int_{\ell} S_{\ell}^* S_{k,\ell}
    \,.
\end{equation}
For the hierarchical region, $S^\sigma_k(p_a,\ell)$ is the soft factor with the graviton $\ell$ being included as a hard leg. Since the cut graviton is still far softer than massive momenta $\ell \ll p_a$, the soft factor can be further expanded in  $\ell$
\begin{align}
    S_k(p_a,\ell)=S_k(p_a) + \delta_1 S_k(p_a,\ell) + \mathcal{O}(\ell^2)
    \,.
\end{align}
Here, $\delta_1 S_k(p_a,\ell)$ denotes terms of $\mathcal{O}(\ell^1)$, and its concrete expression at the leading order is
\begin{align}
    \delta_1 S_k(p_a,\ell) = - \ell\cdot \partial_{p_X} \Stree{-1}{k}|_{p_X=-p_{12}}-\frac{\kappa}{2}\frac{(\epsilon^\sigma_k\cdot\ell)^2}{k\cdot\ell} + \mathcal{O}(\kappa^3)
    \,.
\end{align}
In the classical limit, the $\ell$-independent soft factor can be moved away from the integral by substituting the classical values $p_a=m_a u_a$. Then, we obtain
\begin{align}
i\W^{(1)}_\sigma\Big|_{\rm hierarchical}
=S_k\,P^{(1)}_{\rm abs}
-R\cdot \partial_{p_X} \Stree{-1}{k}
-\frac{\kappa}{2}\!\int_{\ell}\!\frac{(\epsilon^\sigma_k\cdot\ell)^2}{k\cdot\ell}\,\mathcal{N}(\ell) +\mathcal{O}(\kappa^5)\,,
\label{eq:W1hard}
\end{align}
where $P^{(1)}_{\rm abs}, R^{\mu}$, and $\mathcal{N}$ are defined in Sec.~\ref{subsec:inclusive_radiative_matching} and Sec.~\ref{sec:KMOCwaveform}.

Collecting all contributions, the waveform up to the $n=1$ graviton cut is 
\begin{align}
    i\mathcal{W}_{\sigma}&=(|Z|^2+P^{(1)}_{\rm abs})S_k - R\cdot \partial_{p_X} \Stree{-1}{k} +
    |Z|^2 \sum_{\eta} \int_{\ell} S_{\ell}^* S_{k,\ell}-\frac{\kappa}{2}\!\int_{\ell}\!\frac{(\epsilon^\sigma_k\cdot\ell)^2}{k\cdot\ell}\,\mathcal{N}(\ell)
   +\cdots 
   \,.
\end{align}
We expand the soft factors up to the necessary orders for the waveform at $\mathcal{O}(\kappa^3)$ where angular momenta are counted as $\mathcal{O}(\kappa^2)$ because $\mathcal{J}\sim pb \sim Gm^2$. Then, we obtain

\begin{tcolorbox}[colframe=black!75,
  colback=black!3,
  colbacktitle=black!75,
  coltitle=white,
  fonttitle=\bfseries, title = Merger Kernel at NLO, fontupper=\normalsize]
\begin{equation}
\begin{split}
i\mathcal{W}_{\sigma}&=\underbrace{\Stree{-1}{k}+\Stree{0}{k} }_{\text{tree}} \\
    &\underbrace{- R\cdot \partial_{p_X} \Stree{-1}{k}-\frac{\kappa}{2}\!\int_{\ell}\!\frac{(\epsilon^\sigma_k\cdot\ell)^2}{k\cdot\ell}\,\mathcal{N}(\ell)+\Sloop{0}{k}+ \sum_{\eta} \int_{\ell} [\Stree{-1}{\ell}]^* \Stree{-1}{k,\ell} }_{\text{one-loop}~+~(n=1)\text{-cut}}
    \,.
\end{split}
\label{W_soft}
\end{equation}
\end{tcolorbox}
\noindent Here, we have used the matching condition \eqref{eq:matchingZ} together with $|Z|^2=1+\mathcal{O}(\kappa^2)$ and the fact that $\int_{\ell} [\Stree{-1}{\ell}]^* \Stree{-2}{k,\ell}$ vanishes because it is scaleless. Each term of \eqref{W_soft} admits a clear physical interpretation. The first line is the well-known correspondence between the waveform and the soft factor~\cite{Strominger:2014pwa}, but it is now shown to hold only up to the tree-level order. On the second line, the first two terms contribute to the waveform at $\mathcal{O}(\omega^{-1})$. In particular, the first term represents a recoil effect; combined with the Weinberg soft factor on the first line $\Stree{-1}{k}$, the first term shifts the final momentum of the black hole to be $p_X=-p_{12}-R$, correctly describing the radiation loss by $R^{\mu}$,
\begin{align}
    \Stree{-1}{k}\Big|_{p_X=-p_{12}} - R\cdot \partial_{p_X}\Stree{-1}{k}\Big|_{p_X=-p_{12}} = \Stree{-1}{k}\Big|_{p_X=-p_{12}-R} +\mathcal{O}(\kappa^5)
    \,.
\end{align}
The second term, on the other hand, represents the
Christodoulou non-linear memory~\cite{Christodoulou:1991cr,Blanchet:1992br,Thorne:1992sdb}, a memory effect sourced by gravitational wave radiation. The third and fourth terms give logarithmic tails $\omega^0 \ln \omega^{-1}$. In particular, the fourth term is needed to subtract the additional ``quantum'' contributions of $\Sloop{0}{k}$ found in \cite{Sahoo:2018lxl}, which we will elaborate on in Sec.~\ref{sec:oneloop}. In this way, Eq.~\eqref{W_soft} provides a refined relation between the low-frequency waveforms constrained by the classical soft theorems and the quantum soft factors computed by the amplitudes.

Another feature of \eqref{W_soft} is that it clearly disentangles the contributions coming from universal soft physics $\omega \ll \omega_*$ and the ones from strong-gravity physics at $\omega \sim \omega_*$. The latter effects are packaged into particular integrals of the distributions of the radiation states $\mathcal{N}$. This is a general feature of low-energy effective theories where unresolved short-distance physics is integrated out and can be parameterised by effective operators. In our context, our effective theory is formulated for the observable directly, and the amplitude methods have provided a guideline for mapping unresolved strong-gravity physics to the low-frequency waveform through the inclusive KMOC observable. We can then develop a well-defined effective theory for computing the waveform at low frequencies $\omega \ll \omega_*$ from black hole mergers.

All in all, the use of soft theorems provides two advantages. First, they reorganize the KMOC waveform into contributions with distinct physical origins. Second, they separate the universal soft dependence from the hard merger amplitude. This is particularly useful for incorporating the classical spin of the remnant. The physical Kerr state is described by a coherent-spin state, which resums the discrete spin-$j$ sectors and must be retained in the matching condition. However, once a radiative amplitude is reduced by the soft theorem to a universal operator acting on the lower-point merger amplitude, its dependence on the remnant spin can be expanded directly in powers of the classical spin. The coefficients at any fixed low order in this expansion are then extracted from amplitudes with finite quantum spin, without first constructing and resumming the complete higher-point amplitude for arbitrary $j$. These coefficients should be understood as terms in the classical-spin expansion of the fully resummed coherent-state result, and the freedom of the coefficients, if they exist, can be understood as the freedom of effective couplings for mergers. In the present work, we restrict ourselves to the spin-independent, $\mathcal{O}(S_X^0)$, contribution, while the extension to higher orders in $S_X$ is left for future work.

\section{Scalar amplitude test of the NLO merger waveform}
\label{sec:oneloop}
Having established in Sec.~\ref{sec:KMOC} the general low-frequency organization of the KMOC waveform, we now turn to its explicit evaluation. Equation~\eqref{W_soft} already separates the NLO waveform into two qualitatively different types of contributions. The hierarchical region is sensitive to finite-frequency radiation produced by the merger and is therefore not determined by the low-energy amplitude expansion: its effect is encoded in the inclusive hard matching data \(R^\mu\) and \(\mathcal{N}(\ell)\), which determine the remnant recoil and the Christodoulou non-linear memory, respectively. By contrast, the logarithmic \(\omega^0\ln\omega^{-1}\) sector is controlled entirely by the double-soft region and universal long-range dynamics, and can therefore be calculated perturbatively without further information about the strong-field merger process. The purpose of this section is to perform this calculation explicitly in the spin-independent sector. We construct the one-loop radiative soft factor \(S^{(0)}_{k,\mathrm{1\mbox{-}loop}}\) from scalar amplitudes, identify the eikonal and massless poles responsible for its logarithmic terms, and evaluate the accompanying double-soft cut

\begin{equation*}
\int_{\ell}
\bigl[S^{(-1)}_{\ell,\mathrm{tree}}\bigr]^*
S^{(-1)}_{k,\ell,\mathrm{tree}} .
\end{equation*}

This computation provides a non-trivial amplitude-level test of the effective merger description developed in Sec.~\ref{sec:KMOC}: it shows explicitly how the additional logarithms of the quantum one-loop soft factor are removed by the inclusive KMOC cut, leaving precisely the universal classical tail. The scalar calculation therefore determines and verifies the spin-independent logarithmic contribution, but it does not compute the non-perturbative matching data \(R^\mu\) or \(\mathcal{N}(\ell)\), nor the corresponding recoil and non-linear-memory coefficients; these remain properties of the hard radiative merger \(S\)-matrix. Likewise, spin-dependent corrections to the universal logarithmic sector lie beyond the explicit calculation carried out here.

One-loop corrections to the soft factor were computed in Ref.~\cite{Sahoo:2018lxl} by evaluating Feynman diagrams one by one with a cutoff regularisation, where the gravitational drag \eqref{eq:SenMergerDrag} and the early-time acceleration \eqref{eq:SenMergerAcc} are found. Our approach differs in two respects. First, in Sec.~\ref{sec:one-loop}, we reduce the full amplitude to a basis of scalar integrals via generalised unitarity: since the logarithms arise solely from threshold singularities, we can immediately conclude that only the topologies shown in Figures~\ref{fig:topology-D} and \ref{fig:topology-A} can contribute. Second, in Sec.~\ref{sec:log}, we use dimensional regularisation so that the logarithms are read off from reduced master integrals by the method of regions. In Sec.~\ref{sec:cancel}, we reproduce the quantum soft factor, including the additional ``quantum'' logarithms~\eqref{Sgr_quantum}, found in Ref.~\cite{Sahoo:2018lxl}, and show that the ``quantum'' terms are precisely cancelled by the double-soft region integral, leaving only the classical tail. Sec.~\ref{sec:NLOwaveform} assembles all contributions into the complete $\mathcal{O}(\kappa^3)$ waveform, and Sec.~\ref{sec:universal_log} generalizes our analysis to higher orders.

\subsection{Relevant topologies of one-loop four-point amplitude}
\label{sec:one-loop}

The one-loop amplitude is expanded in terms of scalar integrals $I^{(T)}_i$ and analytic terms:
\begin{equation}
\Amp^{\text{1-loop}}_4
=\sum_{T}\sum_{i}c^{(T)}_i I^{(T)}_{i}+ (\text{analytic})\,.
 \label{eq:loopdecomp}
\end{equation}
We are interested in the non-analytic dependence on \(\omega\) arising in the one-loop-corrected soft factor for scalar particles where $T$ denotes topologies and $i$ denotes the arrangement of internal particles. 
Since the leading $\omega^{-1}$ pole is universal according to Weinberg soft theorem, our interest is $\ln \omega^{-1}$ terms of the one-loop amplitude in the soft limit. They arise from threshold singularities: the relevant topologies are those in which the internal lines go on-shell as $\omega \to 0$.

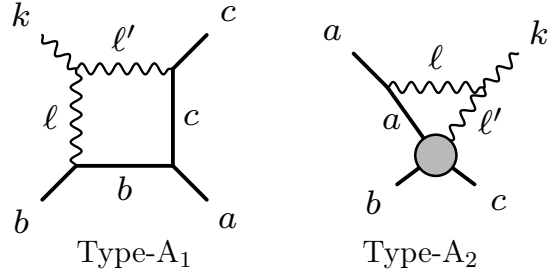
\begin{figure}[t]
\centering
\resizebox{0.5\textwidth}{!}{%
\begin{tabular}{ccccccc}
\scalebox{1.5}{\begin{tikzpicture}[cutdiag]
\coordinate (C) at (0,0); \coordinate (A) at (0,1);
\coordinate (B) at (1,1); \coordinate (D) at (1,0);
\draw[gr] (C) -- node[lab, left=4pt] {$\ell$} (A);
\draw[gr] (A) -- node[lab, above=4pt] {$\ell'$} (B);
\draw[sc] (C) -- node[lab, below=2pt] {$b$} (D);
\draw[sc] (D) -- node[lab, right=2pt] {$c$} (B);
\extleg[gr]{A}{-0.35,0.35}{$k$}
\extleg{C}{-0.35,-0.35}{$b$}
\extleg{B}{0.35,0.35}{$c$}
\extleg{D}{0.35,-0.35}{$a$}
\end{tikzpicture}}
~~&~~
\scalebox{1.5}{\begin{tikzpicture}[cutdiag]
\coordinate (A) at (0,1); \coordinate (B) at (1,1); \coordinate (V) at (0.5,0.3);
\draw[gr] (A) -- node[lab, above=4pt] {$\ell$} (B);
\draw[gr] (B) -- node[lab, right=4pt] {$\ell'$} (V);
\draw[sc] (A) -- node[lab, left=2pt] {$a$} (V);
\extleg{A}{-0.35,0.35}{$a$}
\extleg[gr]{B}{0.35,0.35}{$k$}
\extleg{V}{-0.4,-0.3}{$b$}
\extleg{V}{0.4,-0.3}{$c$}
\node[blob] at (V) {};
\end{tikzpicture}}
\\[2pt]
Type-A$_1$ & Type-A$_2$
\end{tabular}}
\caption{Topologies of the one-loop four-point amplitude that contain two internal graviton lines and exhibit threshold singularities relevant to gravitational drag~\eqref{eq:SenMergerDrag}. Thick lines represent massive states with generic labels $(a,b,c)$, while wavy lines represent gravitons with internal momenta $\ell$ and $\ell'$. In the limit $\omega \to 0$, either the pair $(\ell,c)$ or $(\ell',b)$ goes on shell in type~$\mathrm{A}_1$, whereas the pair $(a,\ell')$ goes on shell in type~$\mathrm{A}_2$. The resulting threshold singularities give rise to the logarithmic terms.
}
\label{fig:topology-D}
\end{figure}

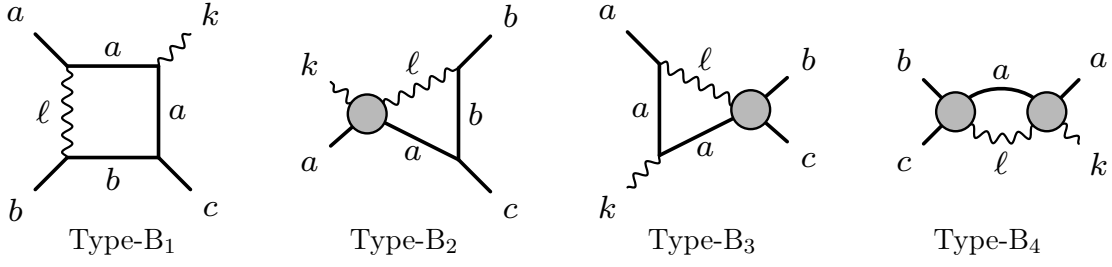
\begin{figure}[t]
\centering
\resizebox{\textwidth}{!}{%
\begin{tabular}{ccccccc}

\scalebox{1.5}{\begin{tikzpicture}[cutdiag]
\coordinate (C) at (0,0); \coordinate (A) at (0,1);
\coordinate (B) at (1,1); \coordinate (D) at (1,0);
\draw[gr] (C) -- node[lab, left=4pt] {$\ell$} (A);
\draw[sc] (A) -- node[lab, above=2pt] {$a$} (B);
\draw[sc] (B) -- node[lab, right=2pt] {$a$} (D);
\draw[sc] (C) -- node[lab, below=2pt] {$b$} (D);
\extleg{A}{-0.35,0.35}{$a$}
\extleg[gr]{B}{0.35,0.35}{$k$}
\extleg{C}{-0.35,-0.35}{$b$}
\extleg{D}{0.35,-0.35}{$c$}
\end{tikzpicture}}
~~&~~

\scalebox{1.5}{\begin{tikzpicture}[cutdiag]
\coordinate (V) at (0,0.5); \coordinate (B) at (1,1); \coordinate (D) at (1,0);
\draw[gr] (V) -- node[lab, above=4pt] {$\ell$} (B);
\draw[sc] (V) -- node[lab, below=2pt] {$a$} (D);
\draw[sc] (B) -- node[lab, right=2pt] {$b$} (D);
\extleg[gr]{V}{-0.4,0.35}{$k$}
\extleg{V}{-0.4,-0.35}{$a$}
\extleg{B}{0.35,0.35}{$b$}
\extleg{D}{0.35,-0.35}{$c$}
\node[blob] at (V) {};
\end{tikzpicture}}
~~&~~

\scalebox{1.5}{\begin{tikzpicture}[cutdiag]
\coordinate (A) at (0,1); \coordinate (C) at (0,0); \coordinate (V) at (1,0.5);
\draw[gr] (A) -- node[lab, above=4pt] {$\ell$} (V);
\draw[sc] (A) -- node[lab, left=2pt] {$a$} (C);
\draw[sc] (C) -- node[lab, below=2pt] {$a$} (V);
\extleg{A}{-0.35,0.35}{$a$}
\extleg[gr]{C}{-0.35,-0.35}{$k$}
\extleg{V}{0.4,0.35}{$b$}
\extleg{V}{0.4,-0.35}{$c$}
\node[blob] at (V) {};
\end{tikzpicture}}
~~&~~

\scalebox{1.5}{\begin{tikzpicture}[cutdiag]
\coordinate (L) at (0,0.5); \coordinate (R) at (1,0.5);
\draw[sc] (L) to[bend left=60] node[lab, above=2pt] {$a$} (R);
\draw[gr] (L) to[bend right=60] node[lab, below=4pt] {$\ell$} (R);
\extleg{R}{0.35,0.35}{$a$}
\extleg[gr]{R}{0.35,-0.35}{$k$}
\extleg{L}{-0.35,0.35}{$b$}
\extleg{L}{-0.35,-0.35}{$c$}
\node[blob] at (L) {};
\node[blob] at (R) {};
\end{tikzpicture}}
\\[2pt]
Type-B$_1$ & Type-B$_2$ & Type-B$_3$  & Type-B$_4$
\end{tabular}
}
\caption{Topologies of the one-loop four-point amplitude with threshold singularities that include only one graviton internal line. The first three are relevant to late-time (early-time) acceleration~\eqref{eq:SenMergerAcc}, while the last one vanishes because, under a BCFW deformation of the cut massive scalar and graviton momenta, the boundary term is absent for the non-derivative $\phi^3$ interaction.}
\label{fig:topology-A}
\end{figure}

We collect these topologies in Figures~\ref{fig:topology-D} and~\ref{fig:topology-A}, where massive lines carry the mutually exclusive labels $(a,b,c)$ and $k$ denotes the observed graviton. The former contains two graviton internal lines, while the latter involves one graviton internal line. Types ${\rm A}_1$, ${\rm A}_2$, ${\rm B}_3$, and ${\rm B}_4$ are symmetric under $b\leftrightarrow c$, so the diagram is uniquely fixed by choosing particle $a$. On the other hand, types ${\rm B}_1$ and ${\rm B}_2$ require specifying two particles, which we choose as $a$ and $b$. Hence, the one-loop amplitude is organised as
\begin{align}
    \Amp^{\text{1-loop}}_4=\sum_{T={\rm A}_1, {\rm A}_2, {\rm B}_3, {\rm B}_4}\sum_a c_a^{(T)}I_a^{(T)} + \sum_{T={\rm B}_1, {\rm B}_2}\sum_{a\neq b}c^{(T)}_{ab} I^{(T)}_{ab} +\cdots
\end{align}
with the dimensionally regularised scalar integrals
\begin{align}
    I^{(T)}_i = -i\int \hd^4 \ell \, \frac{1}{\ell^2}\prod_A \frac{1}{\ell_A^2-m_A^2} \,.
\end{align}
Here, $\ell$ labels the loop momentum of a massless internal line, $\ell_A$ and $m_A=(m_1, m_2,m_X,0)$ are the loop momenta and masses of other propagators, and the Feynman $i\varepsilon$ is implicit. The integrals are evaluated in the sense of 
\begin{align}
    \int \hd^4 \ell = \lim_{\epsilon \to 0} \mu^{2\epsilon}\int \hd^{4-2\epsilon}\ell\,.
\end{align}
Other topologies included in the ellipsis are neglected in the following analysis.

\begin{table}[t]
\centering
\begin{tabular}{cccccc}
\toprule
Type  & $c^{(T)}$ & $I^{(T)}$ & $c^{(T)}\times I^{(T)}$ \\
\midrule
${\rm A}_1$  & $\mathcal{O}(\omega^{2})$  & $\mathcal{O}(\omega^{-2})$ & $\mathcal{O}(\omega^{0})$  \\
${\rm A}_2$ & $\mathcal{O}(\omega^{1})$  & $\mathcal{O}(\omega^{-1})$ & $\mathcal{O}(\omega^{0})$  \\
${\rm B}_1$ & $\mathcal{O}(\omega^{0})$  & $\mathcal{O}(\omega^{-1})$ & $\mathcal{O}(\omega^{-1})$ \\
${\rm B}_2$  & $\mathcal{O}(\omega^{-1})$ & $\mathcal{O}(\omega^{0})$  & $\mathcal{O}(\omega^{-1})$ \\
${\rm B}_3$  & $\mathcal{O}(\omega^{0})$  & $\mathcal{O}(\omega^{0})$  & $\mathcal{O}(\omega^{0})$  \\
${\rm B}_4$  & $0$ & $\mathcal{O}(\omega^{0})$  & $0$ \\
\bottomrule
\end{tabular}
\caption{Summary of the leading $\omega$ scaling of the six topologies of Figures~\ref{fig:topology-D} and~\ref{fig:topology-A}.}
\label{tab:topology-summary}
\end{table}

The leading powers of the graviton frequency $\omega$ of the cut coefficients, of the scalar integrals, and of their products are summarised in Table~\ref{tab:topology-summary}.
The coefficients $c^{(T)}_i$ are uniquely determined by cut conditions of internal particles~\cite{Bern:2011qt,Elvang:2013cua}, and their explicit results are summarized in Table~\ref{tab:cut-coefficients} up to the necessary orders in the low-frequency expansion.\footnote{We verified the box and triangle coefficients through both generalised unitarity methods~\cite{Bern:2011qt,Elvang:2013cua} and Feynman diagram computations, while the bubble coefficients were verified only by the latter using the \textit{Mathematica} packages \texttt{FeynCalc}~\cite{Shtabovenko:2020gxv,Shtabovenko:2023idz} and \texttt{FeynGrav}~\cite{Latosh:2025vax}. Note that we corrected the prefactor of the four-point graviton-scalar contact interaction in \texttt{FeynGrav}~\cite{Latosh:2025vax} by a factor $2$.} They are written in terms of the following invariants
\begin{align}
s_{ab}&:=(p_a+p_b)^2\,,\qquad
y_{ab}:=p_a\cdot p_b\,,\qquad
\omega_{a}:=p_a\cdot k\,,\\
\Delta_{ab}&:=m_a^2\omega_b^2+m_b^2\omega_a^2-2y_{ab}\,\omega_a\omega_b\,,\qquad \Delta:=\Delta_{ab}=\Delta_{bc}=\Delta_{ca}\,,
\label{eq:invariants}
\end{align}
and of the gauge-invariant bilinear
\begin{align}
K_{\sigma}^{(ab)}:=\omega_b\,(p_a\!\cdot\!\epsilon_k^{\sigma})-\omega_a\,(p_b\!\cdot\!\epsilon_k^{\sigma})
=
\begin{cases}
    -[k|p_a p_b|k]/\sqrt{8} & (\sigma=+1)\,, \\
    \bra{k}p_a p_b \ket{k}/\sqrt{8} & (\sigma=-1)
    \,,
\end{cases}
\label{eq:Kab}
\end{align}
with the all-ingoing notation adopted.
While $K_{\sigma}^{(ab)}$ depends a priori on the pair, the momentum conservation together with $k^2=0$ and $k\cdot\epsilon_k^{\sigma}=0$ implies
\begin{equation}
\omega_1+\omega_2+\omega_X=0\,,\qquad
(p_1+p_2+p_X)\cdot\epsilon_k^{\sigma}=0\,,
\label{eq:allingoing_id}
\end{equation}
and hence
\begin{equation}
K_{\sigma}^{(12)}=-K_{\sigma}^{(1X)}=K_{\sigma}^{(2X)}\,,
\end{equation}
so that its square is pair-independent. We therefore drop the superscript and write $K_{\sigma}^2:=(K_{\sigma}^{(ab)})^2$. It is useful to recall that, in terms of these invariants, the Weinberg soft factor can be written as
\begin{align}
\Stree{-1}{k}
 = -\frac{\kappa}{2}\sum_{a=1,2,X}
 \frac{\bigl(p_a\cdot\epsilon^\sigma_k\bigr)^2}{\omega_a}
 = \frac{\kappa}{2}\frac{K_\sigma^2}{\omega_1\omega_2\omega_X}\, .
\label{eq:K2soft}
\end{align}

\begin{table}[t]
  \centering
  \renewcommand{\arraystretch}{2.6}
  \setlength{\tabcolsep}{4pt}
  \resizebox{\textwidth}{!}{%
  \begin{tabular}{@{}r@{\;=\;}l@{}}
    \toprule
    \multicolumn{2}{c}{Cut coefficient} \\
    \midrule
    $c^{(\mathrm{A}_1),\mathrm{LO}}_{a}$
      & $-\dfrac{\kappa^{3}K_{\sigma}^{2}}{\Delta^{2}}
         \left[\left(\Delta + y_{bc}\,\omega_{b}\omega_{c}\right)^{2}
         -\left(m_{b}^{2}m_{c}^{2}-y_{bc}^{2}\right)\omega_{b}^{2}\omega_{c}^{2}\right]$ \\
    $c^{(\mathrm{A}_2),\mathrm{LO}}_{a}$
      & $-\dfrac{\kappa^{3}K_{\sigma}^{2}\,\omega_{a}}{4\Delta^{2}}
         \left[2 s_{bc}\Delta
         +s_{bc}^{2}\left(\omega_{b}^{2}+\omega_{c}^{2}\right)
         +\left(m_{b}^{2}-m_{c}^{2}\right)^{2}\omega_{a}^{2}\right]$ \\
    $c^{(\mathrm{B}_1),\mathrm{LO}}_{ab}$
      & $\dfrac{\kappa^{3}K_{\sigma}^{2}
         \left(m_{a}^{2}m_{b}^{2}-2y_{ab}^{2}\right)}{2\Delta^{2}}
         \left[m_{a}^{2}\Delta
         -2\left(m_{a}^{2}m_{b}^{2}-y_{ab}^{2}\right)\omega_{a}^{2}\right]$ \\
    $c^{(\mathrm{B}_1),\mathrm{NLO}}_{ab}$
      & $0$ \\
    $c^{(\mathrm{B}_2),\mathrm{LO}}_{ab}$
      & $\dfrac{1}{2\omega_{a}}\,c^{(\mathrm{B}_1),\mathrm{LO}}_{ab}$ \\
    $c^{(\mathrm{B}_2),\mathrm{NLO}}_{ab}$
      & $-\dfrac{\kappa^{3}K_{\sigma}^{2}}{2\omega_{a}\Delta^{2}}
         \left[m_{a}^{2}\Delta\left(2y_{ab}\,\omega_{b}-m_{b}^{2}\omega_{a}\right)
         +2\left(m_{a}^{2}m_{b}^{2}-2y_{ab}^{2}\right)
          \left(m_{b}^{2}\omega_{a}-y_{ab}\,\omega_{b}\right)\omega_{a}^{2}\right]$ \\
    $c^{(\mathrm{B}_3),\mathrm{LO}}_{a}$
      & $-\dfrac{\kappa^{3}K_{\sigma}^{2}
         \left(m_{b}^{2}-m_{c}^{2}\right) s_{bc}\,\omega_{a}}{4\Delta^{2}}
         \left[\left(m_{b}^{2}-m_{c}^{2}\right)\omega_{a}
         + s_{bc}\left(\omega_{b}-\omega_{c}\right)\right]$ \\
    $c^{(\mathrm{B}_4)}_{a}$
      & $0$ \\
    \bottomrule
  \end{tabular}}
  \caption{Cut coefficients of the one-loop merger amplitude (non-spinning), in the generic labels of Figures~\ref{fig:topology-D} and~\ref{fig:topology-A}; the coefficient for any assignment of $\{1,2,X\}$ to $(a,b,c)$ follows by relabeling. Only the coefficients required at the corresponding order are listed.}
  \label{tab:cut-coefficients}
\end{table}

\subsection{Eikonal and massless poles of the one-loop soft factor}
\label{sec:log}
The scalar integrals are computed by using the method of regions~\cite{Beneke:1997zp,Jantzen:2011nz}, where long-distance contributions arise from the soft region $k,\ell \ll p_a$. We are particularly interested in logarithmic terms, which are associated with IR/UV divergences in the method of regions:
\begin{align}
    \lim_{\epsilon \to 0}\left(\frac{\mu}{\omega_a}\right)^{2\epsilon}\frac{1}{\epsilon} = \frac{1}{\epsilon}+\ln (\mu^2/\omega_a^2)
    \,.
    \label{log_pole}
\end{align}
This makes life easier because different scalar integrals are reduced to computing IR/UV divergences of simpler master integrals.

Let us explain it by using type ${\rm A}_1, {\rm A}_2$ and ${\rm B}_2$ diagrams since these topologies will be particularly important for classical waveforms. In the soft region, the integrals up to the necessary orders in the small frequency expansion are 
\begin{align}
I^{({\rm A}_1)}_a\big|_{\text{log}}  &=\int \hd^4\ell \frac{-i}{\ell^2 [\ell^2+2k\cdot \ell][2p_c \cdot (\ell+k)][-2p_b \cdot \ell]} + \mathcal{O}(\omega^{-1}\ln \omega)
\,,
\label{ID1_LOsoft}
\\
I^{({\rm A}_2)}_a \big|_{\text{log}} &=\int \hd^4\ell \frac{-i}{\ell^2[\ell^2+2k\cdot \ell][-2p_a\cdot \ell]} + \mathcal{O}(\omega^0 \ln \omega)
\,,
\label{ID2_LOsoft}
\\
I^{({\rm B}_2)}_{ab}\big|_{\text{log}}  &= \int \hd^4 \ell \frac{-i}{\ell^2  [2p_a \cdot (\ell+k)][-2p_b\cdot \ell]} -2 \int \hd^4 \ell \frac{-i k\cdot \ell}{\ell^2 [2p_a \cdot (\ell+k)]^2[-2p_b \cdot \ell]}+\mathcal{O}(\omega^2 \ln 
\omega)
\nn
&=(1+k\cdot \partial_{p_a})\int \hd^4 \ell \frac{-i}{\ell^2  [2p_a \cdot (\ell+k)][-2p_b\cdot \ell]} +\mathcal{O}(\omega^2 \ln \omega)
 \,.
 \label{IA2_NLOsoft}
\end{align}
Here, $\left.I\right|_{\log}$ denotes the logarithmic part of $I$: only terms containing logarithmic singularities are retained after evaluating the integrals. The soft regions also generate non-logarithmic terms, which we omit because they do not contribute to tails of classical waveforms. Note that the $b\leftrightarrow c$ symmetry of $I^{({\rm A}_1)}_a$ is not manifest, but it is indeed the case by using the reparametrisation of the loop momentum $\ell \to -\ell-k$. The first two possess IR divergences, while the third one possesses a UV divergence. These divergences can be extracted from those of scaleless integrals by taking $\ell \ll k$ or $\ell \gg k$. For type $A_1$, an additional IR divergence arises when another massless propagator becomes soft, i.e., $|\ell + k|\ll |k|$. The UV divergence of a scaleless integral is opposite in sign to the IR divergence, so the former is recast into the latter by multiplying by $-1$. Hence, computing the logarithms of \eqref{ID1_LOsoft}-\eqref{IA2_NLOsoft} is reduced to computing the IR divergences of scaleless integrals. Collecting types ${\rm B}_1$ and ${\rm B}_3$ as well, the logarithmic terms of the scalar integrals are given by
\begin{align}
    I^{({\rm A}_1)}_a \big|_{\text{log}} &= \frac{1}{2\omega_c} J_{bk} +\frac{1}{2\omega_b}J_{ck} + \mathcal{O}(\omega^{-1} \ln \omega)
    \,, \label{ID1_LO} \\
    I^{({\rm A}_2)}_a \big|_{\text{log}} &= J_{ak} + \mathcal{O}(\omega^0 \ln \omega)
    \,, \label{ID2_LO} \\
    I^{({\rm B}_1)}_{ab}\Big|_{\text{log}}&=\frac{1}{2\omega_a} J_{ab}  + \mathcal{O}(\omega^0 \ln \omega)
\,, \\
     I^{({\rm B}_2)}_{ab} \big|_{\text{log}} &= - J_{ab}- (k\cdot \partial_{p_a})J_{ab} + \mathcal{O}(\omega^2\ln \omega)
     \,, \label{IA2_NLO}
     \\
     I^{({\rm B}_3)}_a \Big|_{\text{log}} &= J_{a} + \mathcal{O}(\omega^1 \ln \omega)
\,,
\end{align}
with
\begin{align}
    J_{ak}&:= \intIR  \frac{-i}{\ell^2 [2k\cdot \ell][-2p_a\cdot \ell]}
    \,, \label{Jak_def} \\
    J_{ab}&:= \intIR  \frac{-i}{\ell^2 [2p_a\cdot \ell][-2p_b\cdot \ell]} \,,
    \label{Jab_def}
    \\
     J_a&:=\intIR  \frac{+i}{\ell^2 [2p_a\cdot \ell][2p_a\cdot \ell]}
    \,. 
    \label{eq:Ja_def}
\end{align}
The sign of each denominator is defined so that it is replaced by $[\cdots] \to [\cdots +i\varepsilon]$ when the Feynman $i\varepsilon$ is reintroduced. 
Here, ``IR-log'' stands for extracting logarithms from IR poles of the integral by means of \eqref{log_pole}, i.e., $\frac{1}{2\epsilon} \to \ln \omega^{-1}$.\footnote{Precisely speaking, the real part of $J_{ak}$ also contains $\ln^2\omega^{-1}$ terms due to a collinear divergence, which is neglected in Ref.~\cite{Sahoo:2018lxl} where a hard cutoff regularisation is adopted. However, the real part will be cancelled out by \eqref{eq:treeintegrand} and does not contribute to the waveform. Thus, we also only focus on $\ln \omega^{-1}$ arising from the soft divergence by following \cite{Sahoo:2018lxl}.} As we will discuss shortly, these integrals yield $J \sim \omega^0 \ln \omega^{-1}$. Note that we do not distinguish which Lorentz-invariant variable $\omega_a$ appears in the argument of logarithms because the difference between them only affects constant terms.\footnote{As we have explained after \eqref{eq:SenClassical}, the distinction is important to correctly identify whether $\ln \omega^{-1}$ contributes to early-time or late-time waveforms. Using the sign symbol \eqref{eq:varsigmaDef}, one can recover the invariant variable $\omega \to -\varsigma_a \omega_a>0$ by recalling which kinematical threshold is associated with $\ln \omega^{-1}$ and then replace $\omega_a \to \omega_a+i\varepsilon$ according to the Feynman $i\varepsilon$.}

Setting the three-point coupling $\Amp_3$ to unity, we therefore obtain the logarithm of the one-loop amplitude, and hence the one-loop correction to the soft factor, as
\begin{align}
    \Amp_4^{\text{1-loop}}\Big|_{\text{log}}=\Sloop{0}{k}\Big|_{\text{log}}= \sum_m C_m J_m\,, \qquad
m\in\{ab,\,a,\,ak\}\,,
\label{eq:CJdecomp}
\end{align}
with the coefficients
\begin{align}
    C_{ab}&:= 
    \frac{c^{(\text{B}_1), \text{LO}}_{ab}}{2\omega_a} 
    - c^{(\text{B}_2), \text{LO}}_{ab} (1+k\cdot \partial_{p_a}) 
    - c_{ab}^{(\text{B}_2), \text{NLO}} 
    + (a\leftrightarrow b)
    \,, \\
    C_a &:= c^{(\text{B}_3),\text{LO}}_{a}
    \,, \\
    C_{ak}&:= \frac{c^{(\text{A}_1),\text{LO}}_b}{2\omega_c} + \frac{c^{(\text{A}_1),\text{LO}}_c}{2\omega_b} +c^{(\text{A}_2), \text{LO}}_a
    \,.
\end{align}
We note that the type-$\mathrm{B}_1$ and type-$\mathrm{B}_2$ contributions individually contain superclassical terms of order $\mathcal{O}(\omega^{-1}\ln\omega)$. These terms, however, cancel in their sum, 
\begin{align}
    C_{ab}=-c_{ab}^{({\rm B}_2),{\rm LO}} k \cdot \partial_{p_a} - c_{ab}^{(\text{B}_2), \text{NLO}}  + (a\leftrightarrow b)
\end{align}
leaving the final observable free of such contributions. Hence, the coefficients $C_{m}$ are all classical orders $C_m = \mathcal{O}(\omega^0)$.

\subsection{Inclusive cancellation from the double-soft cut}
\label{sec:cancel}

We are in a position to show the cancellation between a certain part of the logarithmic soft factor and the double soft-integral in \eqref{W_soft}. The key observation is that the reduced scalar integrals $J$ can be split into two types of contributions --- the \textit{eikonal pole} and the \textit{massless pole} --- where the first contributions yield the classical waveform \eqref{eq:SenClassical} and the second ones are reduced to the same on-shell integral measure as the double-soft integral, cancelling each other.

The $\ell^0$ integrations in Eqs.~\eqref{Jak_def}, \eqref{Jab_def}, and \eqref{eq:Ja_def} are performed by closing the integration contour around the poles of the propagators, as detailed in Ref.~\cite{Sahoo:2018lxl}. There are two types of poles in the complex $\ell^0$ plane: the eikonal poles $p_a\cdot \ell=0$ and $k \cdot \ell=0$, arising from the linearized propagators, and the massless poles $\ell^2=0$, arising from the massless propagators. The two eikonal poles pinch the $\ell^0$ integration contour, so their contribution is equivalent to taking the unitarity cut of the two linearized propagators, namely,
\begin{align}
    \Im J_{ab} &=  \intIR \hat{\delta}^{(+)} (2p_a\cdot\ell) \hat{\delta}^{(+)} (2p_b\cdot \ell) \frac{i}{\ell^2}
    =  \frac{\delta_{1(a}\delta_{b)2}}{8\pi D_{ab}} \,\ln \omega^{-1}\,,\\
    \Im J_{ak} &=  \intIR \hat{\delta}^{(+)} (2k\cdot\ell) \hat{\delta}^{(+)} (2p_a\cdot \ell) \frac{i}{\ell^2}
    =\frac{\delta_{aX}}{16\pi\,(p_a \cdot k)}\,\ln \omega^{-1}\,,\\
    \Im J_a &=0\,.
    \label{eq:J_imag}
\end{align}
Here, we should recall our kinematics where $a=1,2$ are incoming particles, and $a=X$ and $k$ are physically outgoing. 
The eikonal poles account for the imaginary parts of $J$, as predicted by the optical theorem. On the other hand, the second contribution, the massless pole, is reduced to the on-shell integral of $\ell$:
\begin{align}
\re J_{ab} &= - \intIR \,\hdelta^{(+)}(\ell^2) \frac{1}{[2 p_a\cdot
  \ell][-2 p_b\cdot \ell]} 
  =\frac{1}{32\pi^2 D_{ab}}\,
 \ln\frac{y_{ab}+D_{ab}}{y_{ab}-D_{ab}}\,\ln \omega^{-1} \,,
 \label{eq:J_real} \\
    \re J_{ak} &= - \intIR \,\hdelta^{(+)}(\ell^2) \frac{1}{[2k\cdot \ell][-2p_a\cdot \ell]} = \frac{1}{32\pi^2 (p_a \cdot k)}\,
 \ln\frac{m_a^2}{(p_a \cdot \bar k)^2} \,\ln \omega^{-1}\,,\\
 \re J_{a} &=  \intIR \,\hdelta^{(+)}(\ell^2) \frac{1}{[2 p_a\cdot
  \ell][2 p_a\cdot \ell]} 
=\frac{1}{16\pi^2 m_a^2}\,\ln\omega^{-1}\,.
\end{align}
Combined with the cut coefficients shown in Table~\ref{tab:cut-coefficients}, the eikonal-pole contributions recover the classical logarithmic tail~\eqref{eq:SenNLOtarget}; more precisely, $C_{Xk}\operatorname{Im}J_{Xk}$ gives the gravitational drag in Eq.~\eqref{eq:SenMergerDrag}, whereas $C_{12}\operatorname{Im}J_{12}$ gives the early-time acceleration in Eq.~\eqref{eq:SenMergerAcc}. The massless-pole contributions provide the additional terms of the quantum results~\eqref{Sgr_quantum}.

The waveform \eqref{W_soft} consists of contributions from the soft factor and the on-shell $\ell$-integral from the double-soft region. Recall that the $\ell$-integral is given in the sense of dimensional regularisation. The integral is non-vanishing only when the integrand is inhomogeneous in $\ell$. From Eq.~\eqref{eq:doublesoft}, the only inhomogeneous terms are the Compton terms. Therefore, taking $\sigma=+$ as an example, the contribution from the double-soft region at $\mathcal{O}(\kappa^3)$ is

\begin{align}
\sum_{\eta}\int_{\ell} [S_{\ell^{\eta}}]^* S_{k^+,\ell^\eta}
&=\sum_{\eta}\int_{\ell} [\Stree{-1}{\ell^{\eta}}]^* \Stree{-1}{k^+,\ell^\eta} 
\nn
&=-\left(\frac{\kappa}{2}\right)^3 \!\int\!\hat{\dd}^4\ell\;\hat{\delta}^{(+)}(\ell^2)
\nonumber\\
&\times\sum_{a}
\Bigg\{
\frac{[\ell|p_1p_2|\ell]^2}{(2\ell\cdot p_1)(2\ell\cdot p_2)(2\ell\cdot p_X)}\,
\frac{1}{2p_a\cdot(k+\ell)}\,
\frac{[k|p_a|\ell\rangle^4}{(2k\cdot\ell)(2p_a\cdot\ell)(2p_a\cdot k)}
\nonumber\\
&\hspace{2.6em}
+\frac{\langle\ell|p_1p_2|\ell\rangle^2}{(2\ell\cdot p_1)(2\ell\cdot p_2)(2\ell\cdot p_X)}\,
\frac{1}{2p_a\cdot(k+\ell)}\,
\frac{m_a^4\,[k\ell]^4}{(2k\cdot\ell)(2p_a\cdot\ell)(2p_a\cdot k)}
\Bigg\}\,.
\label{eq:treeintegrand}
\end{align}
Logarithmic terms are computed by using the same analysis outlined in Sec.~\ref{sec:log} together with the tensor reduction of the integrand~\cite{Passarino:1978jh}. The results are found to be equal to the opposite sign of the massless-pole contribution from the one-loop soft factor
\begin{equation}
\left[\re \Sloop{0}{k}
+
\sum_{\eta}\int_{\ell}\,[\Stree{-1}{\ell}]^* \Stree{-1}{k,\ell} \right]_{\ln \omega^{-1}}
=0\,.
\label{eq:cancellation}
\end{equation}
Note that the double-soft contribution is purely real. Consequently, the imaginary part of the one-loop waveform remains unaffected.

\subsection{NLO waveform: universal and matched contributions}
\label{sec:NLOwaveform}

We now assemble the low-frequency merger waveform at $\mathcal{O}(\kappa^3)$ as
\begin{equation}
 \boxed{\;
i\W_\sigma^{\mathcal{O}(\kappa^3)}
=\,i\W_\sigma\Big|_{\rm tail}
+
i \W_\sigma \Big|_{\rm recoil}
+\,i\W_\sigma\Big|_{\rm NL}\,.}
\label{eq:NLOwaveform_summary}
\end{equation}
We summarise these three terms in turn. We use
\begin{tikzpicture}[baseline=-2]
\coordinate (R) at (0,0);
\node[blob, fill=white, minimum size=3.2mm] at (R) {};
\end{tikzpicture}
,
\begin{tikzpicture}[baseline=-2]
\coordinate (R) at (0,0);
\node[blob, minimum size=3.2mm] at (R) {};
\node[circle, fill=white, draw, line width=0.6pt, inner sep=0pt, minimum size=1.5mm] at (R) {};
\end{tikzpicture}
and
\begin{tikzpicture}[baseline=-2]
\coordinate (R) at (0,0);
\node[blob,  minimum size=3.2mm] at (R) {};
\end{tikzpicture}
to denote tree, one-loop and full amplitudes, respectively. 

\paragraph{Logarithmic tail.}
The logarithmic tail, comprising gravitational drag and early-time acceleration, arises from the imaginary part of the loop amplitude,
\begin{align}
    i \W_\sigma\Bigl|_{\rm tail} &=
    i\Im \left[\;
\begin{tikzpicture}[cutdiag, baseline={([yshift=-0.6ex]L)}]
\coordinate (L) at (0,0); \coordinate (R) at (2.4,0);
\draw[sc] (L) -- ++(-0.95,0.6) coordinate (i1); \node[lab] at ($(i1)+(-0.18,0)$) {$1$};
\draw[sc] (L) -- ++(-0.95,-0.6) coordinate (i2); \node[lab] at ($(i2)+(-0.18,0)$) {$2$};
\draw[ms] (L) -- (R);
\coordinate (gk) at (1.2,1.3);
\draw[gr] (R) to[bend right=25] (gk);
\node[lab, above] at ($(gk)+(0,0.06)$) {$k,\,\sigma$};
\draw[sc] (R) -- ++(0.95,0.6);
\draw[sc] (R) -- ++(0.95,-0.6);
\node[blob, fill=white, minimum size=4.6mm] at (L) {};
\node[blob, minimum size=4.6mm] at (R) {};
\node[circle, fill=white, draw, line width=0.6pt, inner sep=0pt, minimum size=2.1mm] at (R) {};
\draw[dashed, line width=0.5pt] (1.2,-0.8) -- (1.2,1.5);
\end{tikzpicture}
\;\right]
\nn
&=\frac{i\kappa^3}{64\pi}\frac{K_\sigma^2}{\omega_1\omega_2}
 \left[
 2\ln(\omega+i\varepsilon)^{-1}+\frac{y_{12}(2y_{12}^2-3m_1^2m_2^2)}{D_{12}^{3}}\ln(\omega-i\varepsilon)^{-1}
 \right]\,.
\label{eq:tailresult}
\end{align}
Here, the diagram is drawn in the sense of the KMOC waveform \eqref{eq:KMOCkerneldef} with the dashed line denoting the final-state cut of the in-in observable. For notational simplicity, we suppress the helicity sum over the cut graviton and the integration over the on-shell phase space of $X$. Note that the imaginary part arises only from types $\text{A}_1, \text{A}_2$, and $\text{B}_2$ in Figures~\ref{fig:topology-D} and \ref{fig:topology-A}, concluding that only three topologies are relevant to the logarithmic tails. The real part of the one-loop amplitude is cancelled by the double-soft region of the one-graviton cut
\begin{equation}
\Re\left[\;
\begin{tikzpicture}[cutdiag, baseline={([yshift=-0.6ex]L)}]
\coordinate (L) at (0,0); \coordinate (R) at (2.4,0);
\draw[sc] (L) -- ++(-0.95,0.6) coordinate (i1); \node[lab] at ($(i1)+(-0.18,0)$) {};
\draw[sc] (L) -- ++(-0.95,-0.6) coordinate (i2); \node[lab] at ($(i2)+(-0.18,0)$) {};
\draw[ms] (L) -- (R);
\coordinate (gk) at (1.2,1.3);
\draw[gr] (R) to[bend right=25] (gk);
\draw[sc] (R) -- ++(0.95,0.6);
\draw[sc] (R) -- ++(0.95,-0.6);
\node[blob, fill=white, minimum size=4.6mm] at (L) {};
\node[blob, minimum size=4.6mm] at (R) {};
\node[circle, fill=white, draw, line width=0.6pt, inner sep=0pt, minimum size=2.1mm] at (R) {};
\draw[dashed, line width=0.5pt] (1.2,-0.8) -- (1.2,1.5);
\end{tikzpicture}
\;\right]
+\int_{\ell}\;
\left[
\begin{tikzpicture}[cutdiag, baseline={([yshift=-0.6ex]L)}]
\coordinate (L) at (0,0); \coordinate (T) at (2.0,0);
\draw[sc] (L) -- ++(-0.95,0.6);
\draw[sc] (L) -- ++(-0.95,-0.6);
\draw[ms] (L) -- (T);
\draw[sc] (T) -- ++(1.0,0.62) coordinate (oa); \node[lab] at ($(oa)+(0.18,0)$) {$1$};
\draw[sc] (T) -- ++(1.0,-0.62) coordinate (ob); \node[lab] at ($(ob)+(0.18,0)$) {$2$};
\coordinate (A) at ($(T)!0.55!(oa)$);          
\draw[gr] (L) to[bend left=40] (A);            
\coordinate (gk) at (1.35,1.4);                
\draw[gr] (A) to[bend right=20] (gk);          
\node[lab] at (0.45,0.70) {$\ell$};
\node[blob, fill=white, minimum size=4.6mm] at (L) {};
\node[blob, fill=white, minimum size=2.6mm] at (A) {};
\draw[dashed, line width=0.5pt] (1.35,-0.8) -- (1.35,1.55);
\end{tikzpicture}
\;+ (1\leftrightarrow 2) \;+\;
\begin{tikzpicture}[cutdiag, baseline={([yshift=-0.6ex]L)}]
\coordinate (L) at (0,0); \coordinate (R) at (1.9,0); \coordinate (T) at (2.5,0);
\draw[sc] (L) -- ++(-0.95,0.6);
\draw[sc] (L) -- ++(-0.95,-0.6);
\draw[ms] (L) -- (R);
\draw[ms] (R) -- (T);                          
\draw[sc] (T) -- ++(0.85,0.55);
\draw[sc] (T) -- ++(0.85,-0.55);
\draw[gr] (L) to[bend left=55] (R);            
\coordinate (gk) at (1.2,1.35);                
\draw[gr] (R) to[bend right=20] (gk);          
\node[lab] at (0.45,0.70) {$\ell$};
\node[blob, fill=white, minimum size=4.6mm] at (L) {};
\node[blob, fill=white, minimum size=2.6mm] at (R) {};   
\draw[dashed, line width=0.5pt] (1.2,-0.8) -- (1.2,1.5);
\end{tikzpicture}
\right]
\;=0\,,
\label{eq:real_part_cancellation_diagram}
\end{equation}
which thereby does not contribute to the waveform, reconciling classical and quantum computations of the gravitational logarithmic tail.

\paragraph{Recoil.}
The hierarchical region $k\ll\ell\lesssim\omega_*$ yields backreaction effects due to radiation. In particular, the recoil appears from the diagram in which the observed graviton is attached to the final cut state of the massive particle:
\begin{equation}
i\W_\sigma\Bigl|_{\rm recoil} = 
-\int_{\ell}\; (\ell\cdot \partial_{p_X})
\left[\;
\begin{tikzpicture}[cutdiag, baseline={([yshift=-0.6ex]L)}]
\coordinate (L) at (0,0); \coordinate (R) at (2.4,0);
\coordinate (V) at (1.85,0);                   
\draw[dashed, line width=0.5pt] (1.2,-0.8) -- (1.2,1.6);
\draw[sc] (L) -- ++(-0.95,0.6);
\draw[sc] (L) -- ++(-0.95,-0.6);
\draw[sc] (R) -- ++(0.95,0.6);
\draw[sc] (R) -- ++(0.95,-0.6);
\draw[ms] (L) -- (R);
\node[lab] at (0.45,0.70) {$\ell$};
\coordinate (gk) at (1.2,1.45);                
\node[lab, above] at ($(gk)+(0,0.06)$) {$k,\,\sigma$};
\draw[gr] (V) to[bend right=25] (gk);          
\draw[gr, preaction={draw, white, line width=2.2pt}]
      (L) to[bend left=45] (R);                
\node[blob,  minimum size=4.6mm] at (L) {};
\node[blob,  minimum size=4.6mm] at (R) {};
\end{tikzpicture}\;\right]
= - R\cdot \partial_{p_X} \Stree{-1}{k}\Big|_{p_X=-p_{12}}\,.
\label{eq:recoil_diagram}
\end{equation}
The recoil contribution shifts the Weinberg factor onto the true remnant momentum, accounting for the momentum loss carried away by radiation \eqref{eq:radiated_momentum}.

\paragraph{Non-linear memory.}
The non-linear memory corresponds to the diagram in which the observed graviton is emitted from the finite-frequency cut graviton,
\begin{equation}    
i\W_\sigma\Bigl|_{\rm NL} =
    \int_{\ell}\;\left[
\begin{tikzpicture}[cutdiag, baseline={([yshift=-0.6ex]L)}]
\coordinate (L) at (0,0);        
\coordinate (R) at (2.6,0);      
\coordinate (V) at (1.93,0.61);  
\draw[sc] (L) -- ++(-1.15,0.65) coordinate (i1);
\node[lab, left] at ($(i1)+(-0.08,0)$) {};
\draw[sc] (L) -- ++(-1.15,-0.65) coordinate (i2);
\node[lab, left] at ($(i2)+(-0.08,0)$) {};
\draw[ms] (L) -- (R);
\draw[gr] (L) to[bend left=42] (V);
\node[lab] at (0.45,0.80) {$\ell$};
\draw[gr] (V) to[bend left=17] (R);
\coordinate (gk) at (1.42,1.42);
\draw[gr] (V) to[bend right=20] (gk);
\node[lab, above] at ($(gk)+(0,0.06)$) {$k,\,\sigma$};
\draw[sc] (R) -- ++(0.95,0.55) coordinate (o1);
\node[lab, right] at ($(o1)+(0.08,0)$) {};
\draw[sc] (R) -- ++(0.95,-0.55) coordinate (o2);
\node[lab, right] at ($(o2)+(0.08,0)$) {};
\node[blob, minimum size=4.6mm] at (L) {};
\node[blob, minimum size=4.6mm] at (R) {};
\draw[dashed, line width=0.5pt] (1.3,-0.85) -- (1.3,1.50);
\end{tikzpicture}
\right]
=-\frac{\kappa}{2}\int_{\ell}\frac{(\epsilon^\sigma_k\cdot\ell)^2}{k\cdot\ell}\,\mathcal{N}(\ell)
\;.
\label{eq:NLdiagram}
\end{equation}
It can be further rewritten as
\begin{equation}
i\W_\sigma\Big|_{\rm NL}
=-\frac{\kappa}{2\omega}\int\dd\Omega_{\ell}\,\frac{\dd E}{\dd\Omega_{\ell}}\,
\frac{(\epsilon^{\sigma}_k\cdot \hat{\bm{n}}_{\ell})^2}{1-\hat{\bm{n}}_k\cdot \hat{\bm{n}}_{\ell}}\,,
\qquad
\frac{\dd E}{\dd\Omega_{\ell}}:=\int\frac{|\bm\ell|^{2}\,\dd|\bm\ell|}{16\pi^3}\,\mathcal{N}(\ell)\,,
\label{eq:nonlinearmemory}
\end{equation}
with $k=-\omega(1,\hat{\bm{n}}_k)$ and $\ell=-|\bm{\ell}|(1,\hat{\bm{n}}_\ell)$: this is the Christodoulou non-linear memory~\cite{Christodoulou:1991cr,Blanchet:1992br,Thorne:1992sdb}, sourced by the angular distribution of the radiated energy. Note that it arises at $\mathcal{O}(G^2)$, one order in $G$ earlier than in $2\to2$ scattering~\cite{Georgoudis:2025vkk}, since the radiation flux sourcing it can appear already at $\mathcal{O}(G)$ for black hole mergers.

\subsection{Higher-order universality of the logarithmic tail}
\label{sec:universal_log}

Finally, we discuss the universality of the leading logarithmic terms found in the classical soft theorem~\cite{Laddha:2018myi,Sahoo:2018lxl,Saha:2019tub, Krishna:2023fxg,Agrawal:2023zea,Boschetti:2026gfd}, in particular how it can be understood in the amplitude computations.

We consider $\mathcal{O}(\kappa^5)$ corrections to the leading logarithms $\omega^0 \ln \omega^{-1}$. At this order, there are three sources for leading logarithms in the KMOC waveform \eqref{eq:KMOCkerneldef}. The first is the $Z$ factor of the three-point amplitude, which renormalises the $\mathcal{O}(\kappa^3)$ logarithms \eqref{eq:tailresult} as
\begin{align}
 |Z|^2 \ima \Sloop{0}{k}(p_a) \Big|_{p_X=-p_{12}}
\,.
\label{eq:Zlog}
\end{align}
The second is the logarithmic correction to the soft factor appearing in the hierarchical region of the $n=1$ graviton cut
\begin{align}
\sum_{\eta'}\int_{X,\ell'}|\widetilde\Gamma^{\eta'}_4(\ell')|^2\,
\Sloop{0}{k}(p_a,\ell') 
\,.
\label{n1_hard}
\end{align}
In this subsection, we use $\ell'$ to denote momenta of finite-frequency gravitons and $\ell$ for the soft graviton momentum.
Finally, we should also include the $n=2$ cut $\mathcal{W}^{(2)}_{\sigma}$, which can be split into several regions depending on hierarchies of $k$ and cut-graviton momenta. From the region where one of the cut-graviton momenta is soft while another is finite, we obtain
\begin{align}
    \sum_{\eta',\eta}\int_{X,\ell',\ell}  |\widetilde\Gamma^{\eta'}_4(\ell')|^2[\Stree{-1}{\ell}(p_a,\ell')]^* \Stree{-1}{k,\ell}(p_a,\ell') 
    \,.
\label{n2_soft+hard}
\end{align}
The sum of the second contribution \eqref{n1_hard} and the third contribution \eqref{n2_soft+hard} is
\begin{align}
\sum_{\eta'}\int_{X, \ell'}  |\widetilde\Gamma^{\eta'}_4(\ell')|^2 \left[ \Sloop{0}{k}(p_a,\ell') + \sum_{\eta} \int_{\ell} [\Stree{-1}{\ell}(p_a,\ell')]^* \Stree{-1}{k,\ell}(p_a,\ell') \right]\,.
\label{eq:kappa5_log}
\end{align}
The brackets have the same structure as the last two terms of \eqref{W_soft}, and we could expect to have the cancellation as in \eqref{eq:cancellation}, leaving the imaginary part  \eqref{eq:SenClassical} only. A remarkable feature is that \eqref{eq:SenClassical} is independent of the final-state massless momenta; in particular, it can be written only in terms of the initial massive momenta if the final massive particle is one~\cite{Sahoo:2018lxl,Sahoo:2021ctw}
\begin{align}
\ima \Sloop{0}{k}(p_a,\ell') \Big|_{p_X=-p_{12}-\ell'} = \ima \Sloop{0}{k}(p_a) \Big|_{p_X=-p_{12}}
\,.
\end{align}
The right-hand side agrees with the classical logarithmic terms \eqref{eq:SenClassical} without the hard graviton. The remaining $\ell'$ integral yields the $n=1$ probability $P^{(1)}_{\rm abs}$,
\begin{align}
    \eqref{eq:kappa5_log} = P^{(1)}_{\rm abs}\,\ima \Sloop{0}{k} (p_a)\Big|_{p_X=-p_{12}} 
 \,.
\end{align}
This contribution is cancelled by the renormalisation of the logarithmic terms \eqref{eq:Zlog},
\begin{align}
(|Z|^2+P^{(1)}_{\rm abs})\ima \Sloop{0}{k}(p_a)\Big|_{p_X=-p_{12}} =\ima \Sloop{0}{k}(p_a)\Big|_{p_X=-p_{12}} +\mathcal{O}(\kappa^7)
\,.
\end{align}
As a result, there would be no correction to the logarithmic tail~\eqref{eq:tailresult} from $\mathcal{O}(\kappa^5)$.

The same cancellations are expected to persist to higher orders in $\kappa$. The hard region of $\mathcal{W}^{(n)}_{\sigma}$ and the region of $\mathcal{W}^{(n+1)}_{\sigma}$ where one cut graviton is soft and $n$ gravitons are hard would give
\begin{align}
&\frac{1}{n!}\sum_{\eta'_1,\cdots, \eta_n'}\int_{X, \ell_1',\cdots,\ell'_n}  |\widetilde\Gamma^{\eta'_1,\cdots, \eta_n'}_{3+n}(\ell_1',\cdots,\ell'_n)|^2 
\nn
&\times \left[ \Sloop{0}{k}(p_a,\ell_1',\cdots,\ell'_n) + \sum_{\eta} \int_{\ell} [\Stree{-1}{\ell}(p_a,\ell_1',\cdots,\ell'_n)]^* \Stree{-1}{k,\ell}(p_a,\ell_1',\cdots,\ell'_n) \right]
\nn
&=\frac{1}{n!}\sum_{\eta'_1,\cdots, \eta_n'}\int_{X, \ell_1',\cdots,\ell'_n}  |\widetilde\Gamma^{\eta'_1,\cdots, \eta_n'}_{3+n}(\ell_1',\cdots,\ell'_n)|^2\, \ima \Sloop{0}{k}(p_a,\ell_1',\cdots,\ell'_n)
\nn
&=P^{(n)}_{\rm abs} \ima \Sloop{0}{k}(p_a)\Big|_{p_X=-p_{12}}
\,.
\end{align}
Therefore, collecting all $n$, we may obtain
\begin{align}
    \left(|Z|^2+\sum_{n\geq 1} P^{(n)}_{\rm abs}\right) \ima \Sloop{0}{k}(p_a)\Big|_{p_X=-p_{12}}=
    \ima \Sloop{0}{k}(p_a)\Big|_{p_X=-p_{12}}\,,
    \label{eq:alln_log}
\end{align}
recovering the prediction that the logarithmic tail is entirely fixed by the data of initial massive particles~\cite{Laddha:2018myi,Sahoo:2018lxl,Saha:2019tub,Krishna:2023fxg,Agrawal:2023zea, Boschetti:2026gfd}. However, it should be noted that this does not mean that cut gravitons did not contribute to the logarithmic term in the amplitude calculation. The cut terms are required to find the proper normalisation \eqref{eq:alln_log} and cannot simply be ignored from the outset.

\section{Conclusions and outlook}
\label{sec:conclusion}

In this paper, we have extended the amplitude-based description of black-hole mergers initiated in Ref.~\cite{Aoki:2024boe} to NLO. The central step is to reorganise the KMOC in-in waveform by applying soft theorems to the cut graviton region by region, as in \eqref{eq:softregions}: the double-soft theorem in the region $\ell\sim k$, and the single-soft theorem acting on the radiative amplitude for $k\ll\ell$. This procedure reduces the $\mathcal{O}(\kappa^3)$ waveform to the four contributions in \eqref{W_soft}, whose evaluation yields the complete low-frequency merger waveform \eqref{eq:NLOwaveform_summary}. The result comprises the logarithmic tail \eqref{eq:tailresult}, fixed entirely by the asymptotic data of the incoming black holes and reproducing the classical logarithmic soft theorem of Ref.~\cite{Sahoo:2018lxl}, together with the recoil contribution \eqref{eq:recoil_diagram} and the Christodoulou non-linear memory \eqref{eq:NLdiagram}, both of which probe the inclusive spectrum $\mathcal{N}(\ell)$ up to the merger scale. Most notably, we find the cancellation \eqref{eq:cancellation}: the ``quantum'' logarithms \eqref{Sgr_quantum} in the one-loop soft factor, whose physical interpretation remained unclear in Ref.~\cite{Sahoo:2018lxl}, are cancelled by the double-soft region of the one-graviton cut, leaving only the eikonal logarithms in the classical waveform. The classical waveform therefore cannot be reconstructed from the radiative amplitude alone; rather, the classical soft theorem emerges from its quantum counterpart through the interplay between the loop amplitude and the KMOC cuts.

We close with several directions for future work. First, the cancellation \eqref{eq:cancellation} deserves a closer examination. Recast as an identity, it relates the real part of the one-loop soft factor to an on-shell phase-space integral of tree-level soft factors. This goes beyond ordinary unitarity, which fixes only the imaginary part of a loop amplitude by its cuts, and suggests a novel factorisation of gravitational amplitudes in the soft limit. Conversely, one may reverse the logic. The classicality of the in-in observable therefore \textit{requires} the ``quantum'' logarithms to be cancelled. In the KMOC formalism, unitarity is known to guarantee the cancellation of superclassical terms~\cite{CaronHuot:2023asymptotic,Herderschee:2023subleading,Bini:2023comparison,Elkhidir:2023radiation,Georgoudis:2023inelastic}; there, the cancellation occurs between the terms linear in $T$ and those bilinear in $T$. The present cancellation is of a different nature. For the merger, all contributions originate from the bilinear terms, and the cancellation~\eqref{eq:cancellation} happens between different cuts. Ordinary unitarity constraints provide no relation between these contributions. The classicality of in-in observables may therefore furnish an independent consistency principle for deriving relations among amplitudes. It would be interesting to determine whether such relations can be established on general grounds~\cite{Cristofoli:2021jas}.\par

Second, at higher orders, the same mechanism persists: the
cancellation between loop and cut merely supplies the normalisation
\eqref{eq:alln_log}, so that the universality of the leading
logarithm is not contaminated by higher-order corrections
(Sec.~\ref{sec:universal_log}). The same framework thus provides a
direct means of testing the tower of leading infrared logarithms
conjectured in Refs.~\cite{Sahoo:2020ryf,AtulBhatkar:2020hqz,Boschetti:2025tru,Banerjee:2026keq} and of exploring the regime in which they can be resummed~\cite{Aoki:2026eos}. Together with the leading-order results of Ref.~\cite{Aoki:2024boe}, these developments point towards a systematic amplitude-based expansion of the merger waveform in the low-frequency regime, complementary to numerical relativity.

Third, a natural extension of our analysis is to incorporate spin. Our organisation of the one-loop computation is particularly well suited to this task. In the decomposition \eqref{W_soft}, all loop corrections are encoded in universal soft factors, while only the three-point merger amplitude is resummed into the coherent-spin amplitude. The NLO waveform can therefore be constructed order by order in the classical spin, without requiring either the one-loop merger amplitude to all orders in spin or its resummation. Moreover, the reduction of the one-loop amplitude to a basis of scalar integrals makes manifest which integral topologies are responsible for the different physical effects. Among the topologies shown in Figures~\ref{fig:topology-D} and~\ref{fig:topology-A}, the type-A topologies capture the gravitational drag, whereas the type-$\text{B}_2$ topology accounts for the early-time acceleration of the two incoming black holes. At the level of individual Feynman diagrams, these effects are instead distributed over many contributions. The master-integral basis thus provides a substantially simpler classification and streamlines the bookkeeping required to make the extension to spin tractable. Since the decomposition \eqref{W_soft} remains valid to all orders in the classical spin, an immediate open question is whether the cancellation in \eqref{eq:cancellation} persists in the presence of spin. Establishing this would extend the logarithmic waveform \eqref{eq:tailresult} to merger processes into a spinning black hole.

The fourth outlook is on the contribution from the hierarchical region: the recoil and the non-linear memory. The integrands in \eqref{eq:recoil_diagram} and \eqref{eq:NLdiagram} require the amplitudes at finite frequencies, whereas their loop expansion controls only the low-frequency regime. A first-principles evaluation of these integrals therefore needs non-perturbative inputs, such as black-hole perturbation theory, and the perturbative amplitude methods may serve as a mapping of such non-perturbative data onto low-frequency waveforms. Nevertheless, optimistically, for relativistic black-hole collisions the relevant dimensionless expansion parameter is $GM\omega$, suggesting that the perturbative results remain useful up to frequencies of order $\omega=\mathcal{O}\bigl((GM)^{-1}\bigr)$.\footnote{Comparing the tree-level waveform with the one-loop waveform \eqref{eq:tailresult}, the actual perturbation parameter is $GM\omega$ for the gravitational drag but $GM\omega/v^3$ for the early-time acceleration, with $v$ being the relative velocity. Hence, for small velocities $v\ll 1$, there is an enhancement of loop corrections, as is well known as the Sommerfeld enhancement, requiring a resummation of iterated topologies; see e.g.,~\cite{Aoki:2026eos}.} At $\omega=\omega_*$, which is about $\omega_*\simeq 0.4/GM$ for the $\ell=2$ quadrupole mode of the Schwarzschild black hole, the spectral waveform must have the exponential cutoff due to the ringdown. Therefore, as an approximation, we may use the perturbative waveform in $\omega\lesssim\omega_*$ and replace it with a ringdown waveform in $\omega\gtrsim\omega_*$; in particular, the dominant contribution of the finite-frequency integrals may then be evaluated by putting the UV cutoff at the ringdown frequency,
\begin{align}
R^{\mu}\approx\int^{\omega_*}\dd\Phi(\ell)\,\ell^{\mu}\,\mathcal{N}(\ell)\,,\qquad
i\W_\sigma\Bigl|_{\rm NL}\approx -\frac{\kappa}{2}\int^{\omega_*}\dd\Phi(\ell)\,\frac{(\epsilon^\sigma_k\cdot\ell)^2}{k\cdot\ell}\,\mathcal{N}(\ell)\,,
\label{eq:QNMcutoff}
\end{align}
with the integrand computed perturbatively. In fact, it is known that low-frequency waveforms from the ultra-relativistic head-on collisions of black holes are well approximated by the memory waveform up to the ringdown frequency~\cite{Smarr:1977fy,Cardoso:2002ay,Sperhake:2008ga,Berti:2010ce}. It would be interesting to make a quantitative comparison between numerical-relativity waveforms~\cite{Pollney:2010hs, Mitman:2020pbt, Mitman:2020bjf, Mitman:2024uss} and the perturbative waveforms, for the latter of which corrections in spin and in the loop expansion can be systematically developed through modern amplitude techniques.\par

More broadly, the results of this work suggest that the effective merger \(S\)-matrix can be developed into a systematic low-frequency theory for black-hole coalescence with cutoff $\omega_*\sim (GM)^{-1}$. The strongly coupled merger region is then integrated out and represented by a hierarchy of inclusive on-shell matching observables, while soft theorems and the KMOC construction determine how these data are resolved by radiation with \(\omega\ll \omega_*\). A natural basis for this hard information is provided by asymptotic energy-flow operators \(\mathcal{E}(\Omega)\). The quantities appearing at the order studied here are already the first members of this hierarchy:
\begin{equation}
\frac{{\rm d}E}{{\rm d}\Omega}
=
\big\langle \mathcal{E}(\Omega)\big\rangle_{\rm merger}\,,
\qquad
R^\mu
=
\int {\rm d}\Omega\,
n^\mu(\Omega)\,
\big\langle \mathcal{E}(\Omega)\big\rangle_{\rm merger}\,,
\end{equation}
so that the recoil and non-linear memory probe different moments of the same one-point energy correlator. At higher orders, the natural new matching data are the connected multipoint energy correlators
\begin{equation}
\big\langle
\mathcal{E}(\Omega_1)\cdots
\mathcal{E}(\Omega_n)
\big\rangle_{{\rm merger}}\,\,,
\qquad n\geq 2\,,
\end{equation}
together, for spinning mergers, with angular-momentum and helicity-flow observables and mixed correlators involving \(\mathcal{E}\) and the corresponding flow operators. Correlations of these radiative observables with the remnant mass, momentum and spin should likewise be regarded as part of the hard matching data. A complete formulation of the program therefore requires deriving the KMOC factorization directly for arbitrary radiative final states, establishing a power counting that determines which \(n\)-point flow correlators can first contribute at each order in \(\omega/\omega_*\), extending the construction to coherent spin and radiated angular momentum, and proving the real--virtual cancellations that remove non-classical soft structures beyond the first non-trivial order. In this formulation the black-disk model supplies only the simplest zeroth-moment matching condition for the inclusive capture probability; the detailed merger dynamics is encoded instead in this tower of inclusive energy- and angular-momentum-flow correlators.

Phenomenologically, such a framework would be complementary rather than alternative to complete inspiral--merger--ringdown models. Its natural regime is one in which the hard process is fully relativistic and strongly gravitational, while the measured observable remains parametrically soft,
\begin{equation}
GM\omega\ll 1,
\end{equation}
without requiring a post-Newtonian expansion in velocity or a small mass ratio. Direct plunges, relativistic head-on collisions, highly eccentric or dynamical-capture mergers, and more general strong-field encounters without a long adiabatic inspiral are therefore particularly natural applications, as are intrinsically infrared observables such as memory, recoil-induced soft radiation and logarithmic tails. In this regime numerical relativity or black-hole perturbation theory could be used as a matching calculation rather than as the complete description of the observable: one would extract the lowest energy-flow correlators, remnant data and, as higher accuracy is required, their multipoint correlations, while universal soft dynamics reconstructs the corresponding long-wavelength waveform. The central conjecture of the effective merger \(S\)-matrix program is therefore that, to any fixed order in the low-frequency expansion, only a finite set of such inclusive correlators is resolved. Establishing this finite-information property and determining its range of validity against numerical-relativity waveforms would promote the construction from an amplitude model of merger into a genuine effective theory of strongly coupled gravitational dynamics viewed at long wavelengths.

\acknowledgments
We thank Zvi Bern, Thibault Damour, and Fei Teng for their valuable insights and advice. We are also grateful to the organizers of the conference Theoretical Developments in Gravitational Wave Physics and to the Yukawa Institute for Theoretical Physics for providing a stimulating environment for these discussions. H.~J. acknowledges \textit{Mathematica} packages \texttt{FeynGrav}~\cite{Latosh:2025vax}, \texttt{FeynCalc}~\cite{Shtabovenko:2020gxv,Shtabovenko:2023idz}, \texttt{SpinorHelicity4D}~\cite{AccettulliHuber:2023ldr} and \texttt{MultivariateApart}~\cite{Heller:2021qkz} for the analysis carried out in Sec.~\ref{sec:oneloop}.  H.~J. gratefully acknowledges the hospitality of National Taiwan University, where part of this work was carried out during a visit supported by the Forefront Physics and Mathematics Program to Drive Transformation (FoPM). The work of K.A. was supported by JSPS KAKENHI Grant No.~JP24K17046. The work of H.~J. was also supported by JSPS KAKENHI Grant No.~JP24KJ0902 and the World Premier International Research Center Initiative (WPI), MEXT, Japan. The work of Y-t H and F-y Cheng is supported by Taiwan National Science and Technology Council grant: 115-2112-M-002 -035 -MY3.

\appendix

\section{Classical limit and \texorpdfstring{$\hbar$}{hbar} counting}
\label{sec:hbarApp}

We collect the $\hbar$ bookkeeping used throughout the paper. This is slightly subtle because the spectral waveform itself is not of order $\hbar^0$. Following the convention of Ref.~\cite{Aoki:2024boe}, the formulas are normally written with $\hbar=1$; when $\hbar$ is restored, the quantum couplings appearing in canonically normalized amplitudes scale at fixed classical couplings as
\begin{equation}
 Z\sim \hbar^{-1/2},\qquad
 \kappa_{\rm q}:=\frac{\kappa}{\sqrt{\hbar}}\sim\hbar^{-1/2},
 \qquad \kappa^2=32\pi G=\mathcal O(\hbar^0).
 \label{eq:hbarcouplings}
\end{equation}
Thus the symbol $\kappa$ used elsewhere in the paper, where $\hbar=1$, should be replaced by $\kappa_{\rm q}$ for the purpose of explicit $\hbar$ counting. The same convention was used for the merger three-point amplitude in the derivation around Eq.~\eqref{3pt_delta}.

The hard classical data are held fixed,
\begin{equation}
 p_i^\mu,\quad p_X^\mu,\quad m_i,\quad m_X,\quad b_i^\mu,\quad S_X^\mu
 =\mathcal O(\hbar^0),
 \label{eq:hbarhard}
\end{equation}
whereas momentum mismatches and on-shell radiation momenta are $\hbar$-soft,
\begin{equation}
 q_i^\mu=\hbar\,\bar q_i^\mu,\qquad
 k^\mu=\hbar\,\omega n^\mu,\qquad
 \ell_r^\mu=\hbar\,\bar\ell_r^\mu,
 \qquad \hbar\to0,
 \label{eq:hbarscaling}
\end{equation}
with $n^2=0$ and $\bar q_i$, $\omega n$, and $\bar\ell_r$ held fixed. This is the KMOC scaling of the momentum mismatch and radiative quanta~\cite{Kosower:2018adc,Cristofoli:2021vyo,Aoki:2024boe}. Restoring dimensions in the wavepacket phase gives
\begin{equation}
 \exp\!\left(-\frac{i}{\hbar}b_i\!\cdot q_i\right)
 =e^{-i b_i\cdot\bar q_i},
 \label{eq:hbarphase}
\end{equation}
so the physical impact parameter remains finite. Equivalently, if one insists on retaining the $\hbar=1$ form $e^{-ib_i\cdot q_i}$ while doing the counting, its Fourier-conjugate variable is $b_i/\hbar=\mathcal O(\hbar^{-1})$; throughout this paper we instead regard $b_i$ as the physical classical impact parameter and restore the explicit $1/\hbar$ in the phase when needed.

The coherent-spin limit is taken simultaneously,
\begin{equation}
 \alpha^I,\tilde\alpha_I=\mathcal O(\hbar^{-1/2}),
 \qquad
 S_X^\mu=\frac{\hbar}{2}\tilde\alpha_I[\sigma_X^\mu]^I{}_J\alpha^J
 =\mathcal O(\hbar^0),
 \label{eq:hbarspin}
\end{equation}
so that the Gaussian factors in Eq.~\eqref{3pt_delta} sharpen into delta functions and impose the classical angular-momentum constraints. For the radiative quantities, Eq.~\eqref{eq:hbarscaling} implies $p\cdot k=\mathcal O(\hbar)$. Consequently the leading emission factor contains one inverse soft propagator in addition to the gravitational coupling and scales as~\cite{Aoki:2024boe}
\begin{equation}
 S_k=\mathcal O(\hbar^{-3/2}),
 \qquad
 i\W_\sigma=\mathcal O(\hbar^{-3/2}).
 \label{eq:hbarwaveform}
\end{equation}
This is precisely the classical scaling required by the field reconstruction. Indeed, the massless phase-space measure is $\dd\Phi(k)=\mathcal O(\hbar^2)$, so Eq.~\eqref{hmunu} scales as
\begin{equation}
 h_{\mu\nu}\sim
 \underbrace{\kappa_{\rm q}}_{\hbar^{-1/2}}
 \underbrace{\dd\Phi(k)}_{\hbar^{2}}
 \underbrace{\W_\sigma}_{\hbar^{-3/2}}
 =\mathcal O(\hbar^0).
 \label{eq:hbarfield}
\end{equation}
Thus it is the spacetime field, rather than the unrescaled spectral waveform, which is literally of order $\hbar^0$.

The terms ``classical'', ``superclassical'', and ``quantum-suppressed'' refer to powers relative to the universal $\hbar^{-3/2}$ weight of Eq.~\eqref{eq:hbarwaveform}: a relative $\hbar^0$ term is classical, a relative $\hbar^{-r}$ term with $r>0$ is superclassical, and a relative $\hbar^{r}$ term is quantum-suppressed. In particular, a relative superclassical $\hbar^{-1}$ contribution to the spectral waveform scales absolutely as $\hbar^{-5/2}$. Such terms can occur in individual loop and cut pieces, but they must cancel in the complete KMOC radiative observable; the smoothness of this cancellation is a general feature of inclusive KMOC observables~\cite{Kosower:2018adc,Cristofoli:2021vyo}.

The counting relevant below is summarized as
\begin{center}
\begin{tabular}{c|c|l}
quantity & $\hbar$ scaling & role \\
\hline
$p_i,m_i,b_i,S_X$ & $\hbar^0$ & fixed classical hard data \\
$q_i,k,\ell_r$ & $\hbar$ & mismatch/radiative momenta \\
$\alpha,\tilde\alpha$ & $\hbar^{-1/2}$ & fixed classical spin \\
$Z,\kappa_{\rm q}$ & $\hbar^{-1/2}$ & quantum amplitude couplings \\
$\dd\Phi(k),\dd\Phi(\ell)$ & $\hbar^2$ & massless phase-space measure \\
$S_k,\W_\sigma$ & $\hbar^{-3/2}$ & leading classical radiative scaling \\
$h_{\mu\nu}$ & $\hbar^0$ & classical spacetime waveform \\
relative NLO superclassical term & $\hbar^{-1}$ & cancels between KMOC sectors
\end{tabular}
\end{center}
This last line explains a point that becomes essential at NLO: a relative $\mathcal O(\hbar)$ correction to the wavepacket/localization kernel cannot be dropped if it multiplies a relative superclassical $\mathcal O(\hbar^{-1})$ term, because their product contributes at the classical order. At fixed order in $G$, this also explains the soft-frequency terminology in the main text. For example, the $\omega^{-1}\ln\omega$ pieces encountered in the one-loop amplitude carry one extra inverse soft propagator relative to the classical NLO logarithm; under $k=\hbar\omega n$ this is precisely one additional inverse power of $\hbar$. This should not be confused with the ordinary Weinberg $1/\omega$ pole at leading order, which is already part of the universal classical scaling in Eq.~\eqref{eq:hbarwaveform}.

Finally, the $\hbar$-soft scaling of $k$ should not be confused with the low-frequency expansion. The limit $\hbar\to0$ at fixed classical frequency $\omega$ in Eq.~\eqref{eq:hbarscaling} already sends the momentum of each individual graviton quantum to zero, as emphasized in Ref.~\cite{Aoki:2024boe}; in this sense a quantum soft limit is built into the classical limit. The low-frequency expansion discussed in this paper, however, is an additional expansion of the resulting classical waveform in the classical frequency,
\begin{equation}
 \hbar\to0\quad\text{at fixed }\omega
 \qquad\text{followed by}\qquad
 \omega\to0.
 \label{eq:orderoflimits}
\end{equation}

\section{BCFW construction of the five-point amplitude}
\label{app:BCFW}
While the double-soft graviton theorem \eqref{eq:doublesoft} was established in Refs.~\cite{Cachazo:2015ksa,Saha:2016kjr,Saha:2017yqi,Chakrabarti:2017ltl}, to make the paper self-contained, in this appendix we construct the five-point tree-level scalar amplitude $\Amp^{\sigma,\eta}_5$ by using BCFW recursion relation and show that its pole structure reproduces the double-soft expansion \eqref{eq:doublesoft}. We adopt the all-incoming convention and the normalization $(\kappa/2)^2\Amp_3=1$. The labels $\sigma$ and $\eta$ denote the helicities of the gravitons with momenta $k$ and $\ell$, respectively. The two massless graviton legs allow the amplitude to be constructed recursively using a two-massless-line BCFW deformation~\cite{Britto:2005fq,Bedford:2005yy}. For the $[k,\ell\rangle$ shift, the spinors are deformed as
\begin{equation}
\begin{aligned}
    |\hat{k}] &= |k]+z|\ell]\,,
    &\qquad |\hat{k}\rangle &= |k\rangle\,,
    \\
    |\hat{\ell}\rangle &= |\ell\rangle-z|k\rangle\,,
    &\qquad |\hat{\ell}] &= |\ell]\,.
\end{aligned}
\label{eq:klshift}
\end{equation}
Alternatively, one may use the conjugate $[\ell,k\rangle$ shift. The corresponding large-$z$ scaling of the deformed amplitude is summarized in Table~\ref{tab:bcfw-large-z}~\cite{Arkani-Hamed:2008owk}. For each helicity configuration, at least one of the two shifts yields an $\mathcal{O}(z^{-2})$ falloff and therefore gives rise to a recursion relation without a boundary contribution. In the following, we choose such a shift for each helicity configuration.
\begin{table}[H]
    \centering
    \begin{tabular}{ccc}
        \toprule
        $(\sigma,\eta)$
        & $[k,\ell\rangle$ shift
        & $[\ell,k\rangle$ shift
        \\
        \midrule
        $(+,+)$ & $\mathcal{O}(z^{-2})$ & $\mathcal{O}(z^{-2})$ \\
        $(+,-)$ & $\mathcal{O}(z^{6})$   & $\mathcal{O}(z^{-2})$ \\
        $(-,+)$ & $\mathcal{O}(z^{-2})$  & $\mathcal{O}(z^{6})$   \\
        $(-,-)$ & $\mathcal{O}(z^{-2})$  & $\mathcal{O}(z^{-2})$ \\
        \bottomrule
    \end{tabular}
    \caption{Large-$z$ scaling of the five-point amplitude under the two BCFW shifts. An $\mathcal{O}(z^{-2})$ falloff implies the absence of a boundary contribution.}
    \label{tab:bcfw-large-z}
\end{table}
A $z$-pole arises only when the two shifted gravitons end up on opposite sides of a factorisation channel: when both lie on the same side, the internal momentum is undeformed. The recursion is therefore built solely from the four-point merger amplitude and the minimally coupled three-point, glued across an on-shell massive line. Denoting by $a\in\{1,2,X\}$ the massive leg attached to the three-point, and by A (B) the channel in which the three-point carries $\hat{k}$ ($\hat{\ell}$),
\begin{equation}
\begin{tabular}{cc}
\ChA{a}{b}{c}, & \ChB{a}{b}{c}
\end{tabular}
,
\label{eq:bcfwchannels}
\end{equation}
each helicity configuration is the sum of six terms,
\begin{equation}
\Amp^{\sigma,\eta}_5 = \sum_{a\in\{1,2,X\}}\left(\Amp^{\sigma,\eta}_{\text{A},a}+\Amp^{\sigma,\eta}_{\text{B},a}\right)\,,
\qquad \{b,c\}=\{1,2,X\}\backslash\{a\}\,.
\label{eq:A5sum}
\end{equation}
With $s_{bc}=(p_b+p_c)^2$ and $r=k+\ell$, it is convenient to define the channel denominators
\begin{equation}
\begin{aligned}
    \Delta_a^{\text{A}} &= \langle k|p_a|k]\, (s_{bc}-m_a^2)\, \langle k|p_a\, r\, p_b|\ell]\, \langle k|p_a\, r\, p_c|\ell]\,,\\
    \Delta_a^{\text{B}} &= \langle \ell|p_a|\ell]\, (s_{bc}-m_a^2)\, \langle k|p_b\, r\, p_a|\ell]\, \langle k|p_c\, r\, p_a|\ell]\,,\\
    \tilde{\Delta}_a^{\text{A}} &= \langle k|p_a|k]\, (s_{bc}-m_a^2)\, \langle \ell|p_b\, r\, p_a|k]\, \langle \ell|p_c\, r\, p_a|k]\,,\\
    \tilde{\Delta}_a^{\text{B}} &= \langle \ell|p_a|\ell]\, (s_{bc}-m_a^2)\, \langle \ell|p_a\, r\, p_b|k]\, \langle \ell|p_a\, r\, p_c|k]\,,
\end{aligned}
\label{eq:channeldenominators}
\end{equation}
where the untilded (tilded) denominators belong to the $[k,\ell\rangle$ ($[\ell,k\rangle$) shift. The residues for all helicity configurations are collected in Table~\ref{tab:5pt_summary}.
\begin{table}[H]
\centering
\renewcommand{\arraystretch}{2.6}
\begin{tabular}{ccll}
\toprule
$(\sigma,\eta)$ & shift & \multicolumn{1}{c}{$\Amp_{\text{A},a}^{\sigma\eta}$} & \multicolumn{1}{c}{$\Amp_{\text{B},a}^{\sigma\eta}$} \\
\midrule
$(+,+)$ & $[k,\ell\rangle$ &
$\dfrac{m_a^4\, [k\ell]^2\, [\ell|p_b p_c|\ell]^2}{\Delta_a^{\text{A}}}$ &
$\dfrac{[\ell|p_a\, r\, p_b p_c\, r\, p_a|\ell]^2}{\langle k\ell\rangle^2\, \Delta_a^{\text{B}}}$ \\
$(-,+)$ & $[k,\ell\rangle$ &
$\dfrac{\langle k|p_a|\ell]^4\, [\ell|p_b p_c|\ell]^2}{[k\ell]^2\, \Delta_a^{\text{A}}}$ &
$\dfrac{\langle k|p_a|\ell]^4\, \langle k|p_b p_c|k\rangle^2}{\langle k\ell\rangle^2\, \Delta_a^{\text{B}}}$ \\
$(-,-)$ & $[k,\ell\rangle$ &
$\dfrac{\langle k|p_a\, r\, p_b p_c\, r\, p_a|k\rangle^2}{[k\ell]^2\, \Delta_a^{\text{A}}}$ &
$\dfrac{m_a^4\, \langle k\ell\rangle^2\, \langle k|p_b p_c|k\rangle^2}{\Delta_a^{\text{B}}}$ \\
$(+,-)$ & $[\ell,k\rangle$ &
$\dfrac{\langle \ell|p_a|k]^4\, \langle \ell|p_b p_c|\ell\rangle^2}{\langle k\ell\rangle^2\, \tilde{\Delta}_a^{\text{A}}}$ &
$\dfrac{\langle \ell|p_a|k]^4\, [k|p_b p_c|k]^2}{[k\ell]^2\, \tilde{\Delta}_a^{\text{B}}}$ \\
\bottomrule
\end{tabular}
\caption{The BCFW channels of the five-point amplitude for all helicity configurations, in units of $(\kappa/2)^2\Amp_3 = 1$. The last two rows are the images of the first two under conjugating all brackets together with $k\leftrightarrow\ell$.}
\label{tab:5pt_summary}
\end{table}
The amplitude \eqref{eq:A5sum} contains more poles than the ones used in its construction. On the single-soft poles, $p_a\cdot k\to0$, its residue reproduces the leading double-soft structure $(\epsilon_k\cdot p_a)^2 \Stree{-1}{\ell}$. On the pole $s_{bc}-m_a^2=2p_a\cdot(k+\ell)+2k\cdot\ell\to0$, the type-A and type-B contributions combine into precisely the Compton amplitude \eqref{eq:contacthel}, establishing the factorisation

\begin{equation}
\lim_{s_{bc}\to m_a^2}\left(s_{bc}-m_a^2\right)\Amp_5^{\sigma\eta}
\;=\;
\Amp^{\text{c}}(p_a;k^{\sigma},\ell^{\eta})\times\Amp_3
\;=\;
\begin{tikzpicture}[x=8mm, y=8mm, line join=round, line cap=round,
  baseline={([yshift=-0.6ex]L)}]
\coordinate (L) at (0,0.5);
\coordinate (R) at (2.0,0.5);
\draw[line width=0.9pt] (L) -- (R);
\draw[line width=0.5pt]
  ($($(L)!0.5!(R)$)+(0,0.11)$) -- ($($(L)!0.5!(R)$)+(0,-0.11)$);
\node[font=\scriptsize, inner sep=1pt, above]
  at ($($(L)!0.5!(R)$)+(0,0.13)$) {$p_a{+}k{+}\ell$};
\draw[line width=0.55pt, decorate,
      decoration={snake, amplitude=0.45mm, segment length=1.5mm}]
  (L) -- ++(-0.5,0.6) coordinate (gk);
\node[font=\scriptsize, inner sep=0pt] at ($(gk)!-7pt!(L)$) {$k\quad$};
\draw[line width=0.55pt, decorate,
      decoration={snake, amplitude=0.45mm, segment length=1.5mm}]
  (L) -- ++(-0.78,0) coordinate (gl);
\node[font=\scriptsize, inner sep=0pt] at ($(gl)!-7pt!(L)$) {$\ell\quad$};
\draw[line width=0.9pt] (L) -- ++(-0.55,-0.65) coordinate (pi);
\node[font=\scriptsize, inner sep=0pt] at ($(pi)!-7pt!(L)$) {$a$};
\node[circle, fill=gray!55, draw, line width=0.6pt,
      inner sep=0pt, minimum size=3.6mm] at (L) {};
\draw[line width=0.9pt] (R) -- ++(0.5,0.6) coordinate (pm);
\node[font=\scriptsize, inner sep=0pt] at ($(pm)!-7pt!(R)$) {$b$};
\draw[line width=0.9pt] (R) -- ++(0.5,-0.6) coordinate (pn);
\node[font=\scriptsize, inner sep=0pt] at ($(pn)!-7pt!(R)$) {$c$};
\end{tikzpicture}\,,
\label{eq:comptonchannel}
\end{equation}
although only the merger four-point and the minimally coupled three-point entered the construction.

\bibliographystyle{utphys}
\bibliography{main}
\end{document}